\documentclass[seceq]{ptptex}

\usepackage{graphicx}

\title{A Realistic Formalism for 4N Bound State in \\a Three-Dimensional Yakubovsky Scheme }

\author{ S. Bayegan$^{1,}$\footnote{E-mail: bayegan@khayam.ut.ac.ir},
M. R. Hadizadeh$^{1,}$\footnote{E-mail: hadizade@khayam.ut.ac.ir},
and W. Gl\"{o}ckle$^{2,}$\footnote{E-mail:
Walter.Gloeckle@tp2.ruhr-uni-bochum.de} }

\inst{
$^1$ Department of Physics, University of Tehran, P.O.Box 14395-547,
Tehran, Iran \\
$^2$ Institute for Theoretical Physics II, Ruhr-University Bochum,
\\ D-44780 Bochum, Germany  }

\abst{
A spin-isospin dependent Three-Dimensional formalism based on the
momentum vectors for the four-nucleon bound state is presented. The
four-nucleon Yakubovsky equations with two- and three-nucleon
interactions are formulated as a function of the vector Jacobi
momenta. Our formalism, according to the number of spin-isospin
states that one takes into account, leads to only a strictly finite
number of the coupled three dimensional integral equations to be
solved. The evaluation of the transition and permutation operators
as well as the coordinate transformations due to considering the
continuous angle variables instead of the discrete angular momentum
quantum numbers are less complicated in comparison with partial wave
representation. With respect to partial wave the present formalism
with the smaller number of equations leads to higher dimensionality
of the integral equations. We have concluded that three dimensional
formalism is less cumbersome for considering the three-nucleon
forces. }

\begin{document}
\newcommand{\intsum}{\sum \kern -15pt \int}

\maketitle

\section{Introduction}

The evaluation of the four-nucleon bound state properties because of
the presence of the fourth nucleon is a challenging task both from
the formalism aspect as well as the numerical one
\cite{Hiyama_PRL85}-\cite{Hadizadeh-EPJA36}. The binding energy of
the $\alpha$-particle with both $2+2$ and $3+1$ structures needs
both realistic two- and three-nucleon forces in order to get closer
to the experimental number \cite{Nogga_PRC65}.

In order to introduce the formalism of four-nucleon bound state we choose the three dimensional (3D) approach.
Let us seek the answer to this main question which is our motivation to select this approach.
Why do we use 3D instead of partial wave (PW) approach?
Few-body calculations are traditionally carried out by solving the
relevant equations in a PW basis. After truncation they lead to
coupled equations on angular momentum quantum numbers. A few PWs
often provide qualitative insight, but modern calculations need many
different spin, isospin and angular momentum combinations. On the other hand the 3D
approach replaces the discrete angular momentum quantum numbers
with continuous angle variables and consequently it considers
automatically all PWs. It may be useful to mention that in PW
approach we should sum all PWs to infinite order, but in practice
we truncate the sum to a finite angular momentum number which is
dependent to the energy that we are working. It means that in
higher energies we will need  more PW components to obtain a
convergence, whereas in 3D approach continues angle variables sum
all PW components to infinite order. So the number of equations in
3D representation is energy independent, whereas in PW
representation it is energy dependent. It appears therefore natural to avoid a PW representation completely
and work directly with vector variables.

The motivation for developing this non PW approach is introducing
a direct solution of the integral equations avoiding the very
involved angular momentum algebra occurring for the permutations,
transformations and especially for the three-body forces.

So in contrast to the truncated PW approach, the number of
equations in the non truncated 3D representation is energy
independent. Therefore this non PW method is more efficient and
applicable to the three- and four-nucleon scattering problems
which consider higher energies than the corresponding bound state
problems. To show this efficiency the leading order
\cite{Fachruddin-PRC68} and full formulation
\cite{Bayegan-in-preparation} of 3N scattering with the inclusion
of 3NFs has been done. Certainly the 4N scattering formulation and
calculations are major additional tasks to be carried out.

We have recently applied the spin-isospin dependent 3D approach based on momentum vectors to
the 3N bound state. The Faddeev equations with NN interactions are
formulated successfully as a function of the vector Jacobi momenta,
specifically the magnitudes of the momenta and the angle between
them, as well as the spin-isospin quantum numbers
\cite{Bayegan-FBS,Bayegan-PRC77}. This novel formalism, according
to the number of spin-isospin states that one takes into account,
leads to only strictly finite number of coupled three dimensional
integral equations to be solved. We have shown that this formalism
for both $^{3}H$ and $^{3}He$ bound states yields the same number
of coupled equations which for fully charge dependent case leads
to only 24 coupled equations. As an application the spin-isospin
dependent three dimensional Faddeev integral equations are solved
with Bonn-B potential \cite{Fachruddin-PRC62, Machleidt-ANP19}. Our result for $^{3}H$ binding energy with
the value of $-8.152$ MeV is in good agreement with the
achievements of the other partial wave based methods. The
calculation of $^{3}H$ binding energy with most modern NN
potentials, i.e. AV18 and chiral potentials, is currently
underway. It should be mentioned that the calculation of $^{3}He$
binding energy, because of the presence of coulomb interaction, is
more complicated in comparison to $^{3}H$ case and with the recent achievement considering
coulomb interaction it can be included in the calculation more easily in future.

We have developed in this article the 3D approach for the
four-nucleon bound state studies with the realistic interactions.
The present work lies a formal ground for numerical investigations
which are planned. This will helps us to reach to the full
solution of the four-nucleon bound state in a straightforward
manner. We propose the spin and isospin degrees of freedom to be
implemented into the three dimensional four-nucleon Yakubovsky
equations. We work directly with the vector variables in the
Yakubovsky scheme in the momentum space, which leads to two
coupled sets of a very limited number of equations in the three
vector variables for the amplitudes which greatly simplifies the
calculations without using a partial wave (PW) decomposition.

In the present article we also show that how three-nucleon forces
(3NFs), which we have chosen for example the Tucson-Melbourne
two-pion exchange 3NF for considering the full solution of the
four-nucleon bound system, can be presented in very simpler
formalism in comparison with the PW \cite{Nogga_PhD}. With regard to
this simplifications the study of Tucson-Melburn 3NF effects in the
$^{3}H$ binding energy is also achievable  and currently is
underway.

This paper is organized as follows. In section
\ref{sec:basis_states} we briefly represent the coupled Yakubovsky
equations for the four-nucleon bound state with two- and
three-nucleon interactions. We also introduce the four-nucleon basis
states in a realistic 3D approach. In sections
\ref{sec:YC_without_3NFs} and \ref{sec:YC_with_3NFs} we derive the
coupled Yakubovsky equations in the realistic 3D approach as a
function of the momentum vectors as well as the spin-isospin quantum
numbers with and without the 3NFs, respectively. We add appendix
\ref{appendix:TM} to show how would be less cumbersome in compare
with the PW approach, if we evaluate the 3NF matrix elements for
example for the Tucson-Melbourne $2\pi$-exchange 3NF in the 3D
approach. In section \ref{sec:number_of_YEs} we discuss about the
number of coupled Yakubovsky equations in both the 3D and the PW
approaches. Finally we summarize in section \ref{sec:summary}.

\section{The 4N Basis States in 3D Representation} \label{sec:basis_states}

\begin{figure}
\begin{center}
\includegraphics*[width=4.50cm]{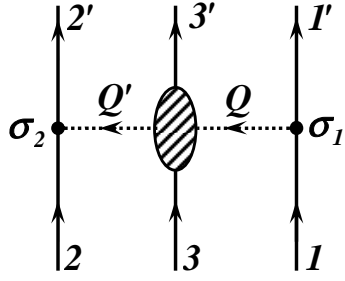}
\end{center}
\caption{\label{fig_3NF} Diagrammatic representation of the part
$V_{123}^{(3)}$ of a $2\pi$ exchange 3NF, like the TM force, and the
definition of the momentum transfers ${\bf Q}$ and ${\bf Q}'$ within
the two subsystems.}
\end{figure}

The bound state of the four nucleon (4N) system, in the presence of
the three-nucleon forces, is described by the two coupled Yakubovsky
equations:
\begin{eqnarray}
\label{YCs} |\psi_{1}\rangle &=& G_{0}tP
[(1-P_{34})|\psi_{1}\rangle+|\psi_{2}\rangle] +(1+G_{0}t) G_{0}
V_{123}^{(3)} |\Psi\rangle, \nonumber \\ |\psi_{2}\rangle &=&
G_{0}t\tilde{P}[(1-P_{34})|\psi_{1}\rangle+|\psi_{2}\rangle],
\end{eqnarray}
where the Yakubovsky components $|\psi_{1}\rangle$ and
$|\psi_{2}\rangle$ belong to $"3+1" (123,4)$ and $"2+2" (12,34)$
partitions of the four particles respectively \cite{Nogga_PRC65}.
Here the free 4N propagator is given by $G_{0}=(E-H_{0})^{-1}$, and
$H_{0}$ stands for the free Hamiltonian. The operator $t$ is the NN
transition matrix determined by a two-body Lippmann-Schwinger
equation. $P$, $\tilde{P}$ and $P_{34}$ are permutation operators.
$P=P_{12}P_{23}+P_{13}P_{23}$ permutes the particles in the
three-body subsystem (123) and $\tilde{P}=P_{13}P_{24}$ interchanges
the two two-body subclusters (12) and (34). The quantity
$V_{123}^{(3)}$ defines a part of the 3NF in the cluster $(123)$,
which is symmetric under the exchange of the particles $1$ and $2$.
This can be related by an interchange of three particles to the two
other parts $V_{123}^{(1)}$ and $V_{123}^{(2)}$ that sum up to the
total 3NF of particles 1,2 and 3:
$V_{123}=V_{123}^{(1)}+V_{123}^{(2)}+V_{123}^{(3)}$. For the 3NFs
based on a meson-exchange picture, $V_{123}^{(3)}$ describes the
interaction induced by a pion interchanged between the particles 1,
2 and on its path rescattered by the third particle, see
Fig.~\ref{fig_3NF}. Applying a combination of the transpositions to
the set of the Yakubovsky components, one obtains the total wave
function $|\Psi\rangle$ as
\begin{equation}
\label{WF}
|\Psi\rangle=[1-(1+P)P_{34}](1+P)|\psi_{1}\rangle+(1+P)(1+\tilde{P})|\psi_{2}\rangle.
\end{equation}

\begin{figure}[hbt]
\begin{center}
\includegraphics*[width=11.0cm]{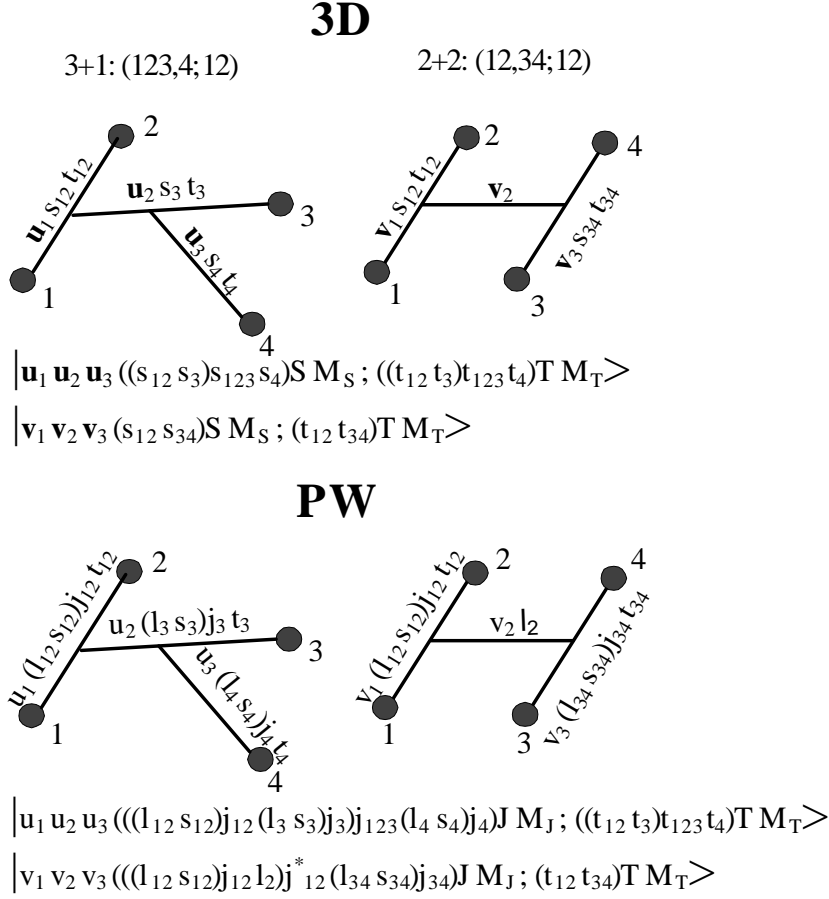}
\end{center}
\caption{\label{fig_basis} Definition of the 4N basis states in both
the 3D and the PW schemes which are constructed by $3+1$ and $2+2$
type of the Jacobi coordinates and the corresponding spin-isospin
quantum numbers.}
\end{figure}

The antisymmetry property of the $|\psi_{1}\rangle$ under the
exchange of the particles $1,2$ and the $|\psi_{2}\rangle$ under
separate exchanges of particles $1,2$ and $3,4$ guarantee that the
$|\Psi\rangle$ is totally antisymmetric. According to the two types
of partitions $(123,4)$ and $(12,34)$, as shown in
Fig.~\ref{fig_basis}, we introduce two different sets of the 4N
basis states in a 3D representation which are suitable to represent
the Yakubovsky components $|\psi_{1}\rangle$ and $|\psi_{2}\rangle$
in the coupled equations (\ref{YCs}):
\begin{eqnarray}
\label{basis}
  |\, {\bf u} \,\,  \alpha  \, \rangle &\equiv&
  | \, {\bf u}_{1}\,{\bf u}_{2}\,{\bf u}_{3} \,\, \alpha_{1234} \,  \rangle,   \nonumber \\
 |\, {\bf v} \,\,  \beta  \, \rangle &\equiv&
 |\, {\bf v}_{1}\,{\bf v}_{2}\,{\bf v}_{3} \,\, \beta_{1234} \,
 \rangle .
\end{eqnarray}
The spin-isospin part of the basis states are:
\begin{eqnarray}
\label{basis_spin-isospin} | \, \alpha_{1234} \,  \rangle &\equiv&
   | \,  \alpha^{S}_{1234} \, \alpha^{T}_{1234}\,  \rangle,  \nonumber \\
 |\,   \beta_{1234}  \, \rangle &\equiv&
  |\,  \beta^{S}_{1234} \, \beta^{T}_{1234}\,
 \rangle,
\end{eqnarray}
where
\begin{eqnarray}
\label{basis_spin} | \, \alpha^{S}_{1234} \, \rangle &=&
 | \, (((s_{1}\,\,s_{2})s_{12} \,\, s_{3})s_{123} \,\, s_{4}) S \, M_{S} \, \rangle
 \equiv
 | \, ((s_{12} \,\, \frac{1}{2})s_{123} \,\, \frac{1}{2}) S \, M_{S} \, \rangle  , \nonumber \\
  | \, \beta^{S}_{1234} \, \rangle  &=&
 | \, ((s_{1}\,\,s_{2})s_{12} \,\, (s_{3}\,\,s_{4})s_{34})   S \, M_{S} \, \rangle
 \equiv
 | \, (s_{12}\,\, s_{34}) S \, M_{S} \, \rangle  ,
\end{eqnarray}
and the isospin parts of the basis states $| \, \alpha^{T}_{1234} \,
\rangle$ and $| \, \beta^{T}_{1234} \, \rangle$ are similar to the
spin parts.

Each one of the basis states involves three standard Jacobi momenta
${\bf u}_{1},\,{\bf u}_{2}$ and ${\bf u}_{3}$ or ${\bf v}_{1},\,{\bf
v}_{2}$ and ${\bf v}_{3}$ \cite{Nogga_PRC65}. As indicated in the
Fig. (\ref{fig_basis}) the angular dependence explicitly appears in
the Jacobi vector variables, whereas in a standard PW representation
the angular dependence leads to three orbital angular momentum
quantum numbers for each kind of the basis states, i.e. $l_{12},
l_{3}$ and $l_{4}$ or $l_{12}, l_{2}$ and $l_{34}$ \cite{Nogga_PhD}.
It indicates that in our 3D formalism there is not any coupling
between the orbital angular momenta and corresponding spin quantum
numbers. Therefore we couple the spin quantum numbers $s_{12}, \,
s_{3}$ and $s_{4}$ or $s_{12}$ and $s_{34}$ to the total spin $S$
and its third component $M_{S}$, by using only one intermediate
quantum number $s_{123}$ for the first basis states: $| \, ((s_{12}
\,\, s_{3})s_{123} \,\, s_{4}) S \, M_{S} \, \rangle$ or $|
\,(s_{12} \,\, s_{34}) S \, M_{S} \, \rangle$. For the isospin
quantum numbers similar coupling schemes to the total isospin $T \,
M_{T}$ involve one intermediate quantum number $t_{123}$: $| \,
((t_{12} \,\, t_{3})t_{123} \,\, t_{4}) T \, M_{T} \, \rangle$ or $|
\,(t_{12} \,\, t_{34}) T \, M_{T} \, \rangle$. In order to be able
to evaluate the transition and permutation operators we need the
free 4N basis states $|\, {\bf u} \,\,  \gamma \, \rangle$ and $|\,
{\bf v} \,\, \gamma \, \rangle$, where
\begin{eqnarray}
\label{spin_free}
 | \, \gamma \, \rangle \equiv | \, \gamma_{1234} \, \rangle \equiv | \, \gamma^{S}_{1234} \, \gamma^{T}_{1234} \,
 \rangle, \,\,\,\,\,
  | \, \gamma^{S}_{1234} \, \rangle \equiv
 | \, m_{s_{1}} \, m_{s_{2}} \, m_{s_{3}} \, m_{s_{4}} \, \rangle,
  \end{eqnarray}
where $m_{s_{i}}, i=1,2,3,4$ indicates the projection of the spin of
each nucleon. The isospin part of the basis states $| \,
\gamma^{T}_{1234} \, \rangle$ is similar to the spin part. In this
respect when we are going from one of the 4N basis states, $| \,
\alpha \, \rangle$ and $| \, \beta \, \rangle$, to the free 4N basis
states, $| \, \gamma \, \rangle$, we should calculate the following
Clebsch-Gordan coefficients;
\begin{eqnarray}
\label{CG} \langle \, \gamma  | \, \alpha \, \rangle &\equiv&
 g_{\gamma \, \alpha } \equiv
 g_{\gamma_{1234} \, \alpha_{1234} }
\equiv g_{\gamma_{1234} \, \alpha_{1234}  }^{S} \, g_{\gamma_{1234}
\, \alpha_{1234}  }^{T} \nonumber
\\ &=& \langle \, m_{s_{1}} \, m_{s_{2}} \, m_{s_{3}} \, m_{s_{4}}
| \, ((s_{12} \,\, \frac{1}{2})s_{123} \,\, \frac{1}{2}) S \,
M_{S} \, \rangle \nonumber \\ &&\times \langle \, m_{t_{1}} \,
m_{t_{2}} \, m_{t_{3}} \, m_{t_{4}} | \, ((t_{12} \,\,
\frac{1}{2})t_{123} \,\, \frac{1}{2}) T \, M_{T}  \, \rangle,
\nonumber \\\nonumber \\ \langle \, \gamma  | \, \beta \, \rangle
&\equiv& g_{\gamma \beta } \equiv g_{\gamma_{1234} \, \beta_{1234}
} \equiv g_{\gamma_{1234} \, \beta_{1234} }^{S} \,
g_{\gamma_{1234} \, \beta_{1234}  }^{T} \nonumber
\\ &=& \langle \, m_{s_{1}} \, m_{s_{2}} \, m_{s_{3}} \, m_{s_{4}}
| \, (s_{12} \,\, s_{34}) S \, M_{S}  \, \rangle  \nonumber \\
&&\times \langle \, m_{t_{1}} \, m_{t_{2}} \, m_{t_{3}} \, m_{t_{4}}
| \, (t_{12} \,\, t_{34}) T \, M_{T}  \, \rangle.
\end{eqnarray}

The introduced basis states are complete in the four-nucleon Hilbert
space as:
\begin{eqnarray}
\label{complete}
 \intsum_{\xi}^{\bf
A}  \,\,|\, {\bf A} \, \xi \,\rangle \,\langle\, {\bf A}\, \xi \,|=
\mathbf{1} , \,\, \intsum_{\xi}^{\bf A}\equiv\sum_{\xi} \int D^{3}A
\equiv\sum_{\xi} \int d^{3}A_{1}\, \int d^{3}A_{2}\, \int
d^{3}A_{3},
\end{eqnarray}
where ${\bf A}$ indicates each one of the ${\bf u}$ and ${\bf v}$
vector sets and $\xi$ indicates $\alpha, \, \beta$ and $\gamma$
quantum number sets. They are also normalized according to:
\begin{equation}
\label{normalization} \langle \, {\bf A}\,\xi\, |\,{\bf A}'\, \xi'\,
\rangle=\delta^{3}({\bf A}-{\bf A}')\,\delta_{\xi\,\xi'}.
\end{equation}

Clearly the basis states $|\, {\bf u} \,\,  \alpha  \, \rangle $ are
adequate to expand the first Yakubovsky component $|\psi_{1}\rangle$
and correspondingly the basis states $|\, {\bf v} \,\,  \beta  \,
\rangle$ are adequate for the second one $|\psi_{2}\rangle$.

\section{Yakubovsky Equations in 3D Representation Without 3NFs} \label{sec:YC_without_3NFs}
Let us now represent the coupled equations (\ref{YCs}), without the
3NFs, with respect to the basis states which have been introduced in
Eq. (\ref{basis}):
\begin{eqnarray}
\label{YCs_basis1}
 \langle \, {\bf u}\, \alpha \,|\psi_{1}\rangle &=&
\intsum_{\alpha''}^{\bf u''} \,\langle \, {\bf u} \, \alpha
\,|G_{0}t P(1-P_{34})| \, {\bf u}'' \, \alpha'' \,\rangle \langle
{\bf u}'' \, \alpha''\, |\psi_{1}\rangle
  \nonumber \\  &+& \intsum_{\beta'}^{\bf v'}  \,\langle  \, {\bf u} \, \alpha \, |G_{0}t P|
 \, {\bf v}' \, \beta' \, \rangle \langle
 \, {\bf v}' \, \beta' \, |\psi_{2}\rangle,
 \nonumber \\ \nonumber \\
\langle  \, {\bf v} \, \beta \, |\psi_{2}\rangle &=&
\intsum_{\alpha'}^{\bf u'} \, \langle  \, {\bf v} \, \beta \,
|G_{0}t \tilde{P}(1-P_{34})|
 \, {\bf u}' \, \alpha' \, \rangle \langle
 \, {\bf u}' \, \alpha' \,|\psi_{1}\rangle
  \nonumber \\  &+& \intsum_{\beta'}^{\bf v'} \,\langle  \, {\bf v} \, \beta \, |G_{0}t \tilde{P}|
 \, {\bf v}' \, \beta' \, \rangle \langle
 \, {\bf v}' \, \beta' \, |\psi_{2}\rangle.
\end{eqnarray}

It is convenient to insert the free 4N completeness relations
between the permutation operators, it results:
\begin{eqnarray}
\label{YCs_basis2}
 \langle \, {\bf u}\, \alpha \, |\psi_{1}\rangle &=&
  \intsum_{\gamma'}^{\bf u'} \, \intsum_{\alpha''}^{\bf u''} \, \langle \, {\bf u}\, \alpha \, |G_{0}t P |  \, {\bf u}'
\, \gamma' \, \rangle  \langle \, {\bf u}' \, \gamma' \, |
(1-P_{34})| \, {\bf u}'' \, \alpha'' \, \rangle  \langle \, {\bf
u}'' \, \alpha'' \, |\psi_{1}\rangle \nonumber
\\  &+&  \intsum_{\gamma'}^{\bf u'} \, \intsum_{\beta'}^{\bf v'}
 \,\ \langle \, {\bf u}\, \alpha \, |G_{0}t P | \, {\bf u}'
\, \gamma' \, \rangle  \langle \, {\bf u}' \, \gamma' \, | \, {\bf
v}' \, \beta' \, \rangle   \langle \, {\bf v}' \, \beta' \,
|\psi_{2}\rangle,
 \nonumber \\ \nonumber \\
\langle \, {\bf v}\, \beta \, |\psi_{2}\rangle &=& \,
\intsum_{\gamma'}^{\bf v'} \, \intsum_{\alpha'}^{\bf u'} \, \langle
\, {\bf v}\, \beta \, |G_{0}t \tilde{P}| \, {\bf v}' \, \gamma' \,
\rangle  \langle \, {\bf v}' \, \gamma' \, | (1-P_{34})| \, {\bf u}'
\, \alpha' \,
 \rangle \langle \, {\bf u}' \, \alpha' \,
|\psi_{1}\rangle
  \nonumber \\  &+& \, \intsum_{\beta'}^{\bf v'} \, \langle \, {\bf v}\, \beta \, |G_{0}t \tilde{P}| \, {\bf
v}' \, \beta' \, \rangle \langle \, {\bf v}' \, \beta' \,
|\psi_{2}\rangle.
\end{eqnarray}

For evaluating the above coupled equations, Eq. (\ref{YCs_basis2}),
we need to evaluate the following matrix elements:
\begin{eqnarray}
 && \langle \, {\bf u}\, \alpha \,
|G_{0}t P | \, {\bf u}' \, \gamma' \, \rangle,
 \label{matrix_elements_1}  \\
&& \langle \, {\bf v}\, \beta \, |G_{0}t \tilde{P}| \, {\bf v}' \,
\gamma' \, \rangle  , \quad  \langle \, {\bf v}\, \beta \, |G_{0}t
\tilde{P}| \, {\bf v}' \, \beta' \, \rangle,
\label{matrix_elements_2}
\\ && \langle \, {\bf u}' \, \gamma' \, |
1-P_{34}| \, {\bf u}'' \, \alpha'' \, \rangle,
\label{matrix_elements_3}  \\ && \langle \, {\bf v}' \, \gamma' \,
| 1-P_{34}| \, {\bf u}' \, \alpha' \, \rangle,
\label{matrix_elements_4} \\ && \langle \, {\bf u}' \, \gamma' \,|
\, {\bf v}' \, \beta' \,  \rangle. \label{matrix_elements_5}
\end{eqnarray}

For evaluating the first term, Eq. (\ref{matrix_elements_1}), we
should insert again a free 4N completeness relation between the NN
$t$-matrix operator and the permutation operator $P$ as:
\begin{eqnarray}
\label{G0_T_P}
 \langle \, {\bf u}\, \alpha \, |G_{0}t P | \, {\bf u}' \,
\gamma' \, \rangle &=& \frac{1}{E-\frac{u_{1}^{2}}{m}
-\frac{3u_{2}^{2}}{4m}-\frac{2u_{3}^{2}}{3m}} \, \sum_{\gamma'''}
\, \intsum_{\gamma''}^{\bf u''} \nonumber \\ &\times& \langle \,
\alpha | \gamma''' \, \rangle \langle \, {\bf u} \, \gamma''' \,
|t| \, {\bf u}'' \, \gamma'' \, \rangle
 \langle \, {\bf u}'' \, \gamma'' \, |
P| \, {\bf u}' \, \gamma' \, \rangle,
\end{eqnarray}
where the matrix elements of the NN $t$-matrix and the permutation
operator $P$ are evaluated separately as:
\begin{eqnarray}
\label{t_in_u} \langle \, {\bf u} \, \gamma''' \, |t | \, {\bf u}''
\, \gamma'' \, \rangle &=& \delta^{3}({\bf u}_{2}-{\bf u}''_{2})\,
\delta^{3}({\bf u}_{3}-{\bf u}''_{3})\, \delta_{m''_{s_{3}}
m'''_{s_{3}}} \, \delta_{m''_{s_{4}} m'''_{s_{4}}} \, \,
\delta_{m''_{t_{3}} m'''_{t_{3}}} \, \delta_{m''_{t_{4}}
m'''_{t_{4}}} \nonumber
\\ &\times&  \, \langle \, {\bf u}_{1} \, m'''_{s_{1}}
m'''_{t_{2}} m'''_{t_{1}} m'''_{s_{2}} |t(\epsilon) |{\bf u}''_{1}
\, m''_{s_{1}} m''_{s_{2}} m''_{t_{1}} m''_{t_{2}} \rangle,
\nonumber
\\  \epsilon &=& E-\frac{3u_{2}^{2}}{4m}-\frac{2u_{3}^{2}}{3m},
\end{eqnarray}
\begin{eqnarray}
\label{P_in_u} \langle \, {\bf u}'' \, \gamma'' \,| P| \, {\bf u}'
\, \gamma' \, \rangle &=& \delta^{3}({\bf u}''_{3}-{\bf u}'_{3}) \,
\delta_{m''_{s_{4}} m'_{s_{4}}} \, \delta_{m''_{t_{4}} m'_{t_{4}}}
 \nonumber
\\ &\times&  \Biggl\{\, \delta^{3}({\bf u}''_{1}+\frac{1}{2}{\bf u}'_{1}+
\frac{3}{4}{\bf u}'_{2}) \, \delta^{3}({\bf u}''_{2} -{\bf
u}'_{1}+ \frac{1}{2}{\bf u}'_{2} ) \nonumber \\ && \quad \times
  \delta_{m''_{s_{1}} m'_{s_{2}}} \, \delta_{m''_{s_{2}} m'_{s_{3}}} \, \delta_{m''_{s_{3}} m'_{s_{1}}}
 \, \delta_{m''_{t_{1}} m'_{t_{2}}} \, \delta_{m''_{t_{2}}
m'_{t_{3}}} \, \delta_{m''_{t_{3}} m'_{t_{1}}}
 \nonumber \\ &&  \,\, +
  \delta^{3}({\bf u}''_{1}+\frac{1}{2}{\bf u}'_{1}-
\frac{3}{4}{\bf u}'_{2}) \, \delta^{3}({\bf u}''_{2} +{\bf
u}'_{1}+ \frac{1}{2}{\bf u}'_{2} ) \nonumber \\ && \quad \times
\delta_{m''_{s_{1}} m'_{s_{3}}} \, \delta_{m''_{s_{2}} m'_{s_{1}}}
\, \delta_{m''_{s_{3}} m'_{s_{2}}} \, \delta_{m''_{t_{1}}
m'_{t_{3}}} \, \delta_{m''_{t_{2}} m'_{t_{1}}} \,
\delta_{m''_{t_{3}} m'_{t_{2}}}
 \, \Biggr\}. \nonumber \\
\end{eqnarray}

For evaluation of the matrix elements of the permutation operator
$P$ we have used the relation between the Jacobi momenta in
different three-body subsystems $(312,4),(231,4)$ and $(123,4)$.
Inserting Eqs. (\ref{t_in_u}) and $(\ref{P_in_u})$ into Eq.
(\ref{G0_T_P}) leads to:
\begin{eqnarray}
\label{G0_T_P_in_u} \langle \, {\bf u}\, \alpha \,  |G_{0}t P | \,
{\bf u}' \, \gamma' \, \rangle &=& \frac{\delta^{3}({\bf
u}_{3}-{\bf u}'_{3})}{E-\frac{u_{1}^{2}}{m}
-\frac{3u_{2}^{2}}{4m}-\frac{2u_{3}^{2}}{3m}} \, \sum_{\gamma'''}
\, g_{\alpha \gamma'''} \, \delta_{m'''_{s_{4}} m'_{s_{4}}} \,
\delta_{m'''_{t_{4}} m'_{t_{4}}}  \nonumber \\ &\times& \Biggl \{
\, \delta^{3}({\bf u}_{2} -{\bf u}'_{1}+ \frac{1}{2}{\bf u}'_{2}
)\, \delta_{m'''_{s_{3}} m'_{s_{1}}} \, \delta_{m'''_{t_{3}}
m'_{t_{1}}} \nonumber \\ && \quad \times \langle {\bf u}_{1} \,
m'''_{s_{1}} m'''_{s_{2}} \, m'''_{t_{1}} m'''_{t_{2}}
|t(\epsilon) |\frac{-1}{2}{\bf u}_{2} -{\bf u}'_{2} \, m'_{s_{2}}
m'_{s_{3}} \, m'_{t_{2}} m'_{t_{3}} \rangle \nonumber \\ && \,\, +
\delta^{3}({\bf u}_{2} +{\bf u}'_{1}+ \frac{1}{2}{\bf u}'_{2} )\,
\delta_{m'''_{s_{3}} m'_{s_{2}}} \, \delta_{m'''_{t_{3}}
m'_{t_{2}}} \nonumber \\ && \quad \times \langle{\bf u}_{1} \,
m'''_{s_{1}} m'''_{s_{2}} \, m'''_{t_{1}} m'''_{t_{2}}
|t(\epsilon) |\frac{1}{2}{\bf u}_{2} +{\bf u}'_{2} \, m'_{s_{3}}
m'_{s_{1}} \, m'_{t_{3}} m'_{t_{1}} \rangle \, \Biggr\}. \nonumber
\\
\end{eqnarray}

Representation of the second terms, Eq. (\ref{matrix_elements_2}),
follows the similar steps:
\begin{eqnarray}
 \label{G0_T_Pbar_1}
\langle{\bf v} \, \beta \, |G_{0}t \tilde{P} |{\bf v}' \, \beta' \,
\rangle &=& \sum_{\gamma'} \, \, \langle{\bf v} \, \beta \, |G_{0}t
\tilde{P} |{\bf v}' \, \gamma' \, \rangle \, \langle \, \gamma' |
\beta' \, \rangle \equiv  \sum_{\gamma'} \, g_{\gamma' \beta' } \,
\langle{\bf v} \, \beta \, |G_{0}t \tilde{P} |{\bf v}' \, \gamma' \,
\rangle  , \nonumber \\
\end{eqnarray}
where
\begin{eqnarray}
 \label{G0_T_Pbar}
 \langle{\bf v} \, \beta \,
|G_{0}t \tilde{P} |{\bf v}' \, \gamma' \, \rangle &=&
\frac{1}{E-\frac{v_{1}^{2}}{m}
-\frac{v_{2}^{2}}{2m}-\frac{v_{3}^{2}}{m}} \, \sum_{\gamma'''} \,
\, \intsum_{\gamma''}^{\bf v''} \nonumber \\ &\times&  \langle \,
\beta | \gamma''' \, \rangle \, \langle \, {\bf v} \, \gamma''' \,
|t | \, {\bf v}'' \, \gamma'' \, \rangle \, \langle \, {\bf v}''
\, \gamma'' \, | \tilde{P}| \, {\bf v}' \, \gamma' \, \rangle.
\end{eqnarray}

The matrix elements of the NN $t$-matrix and the permutation
operator $\tilde{P}$ are evaluated as:
\begin{eqnarray}
\label{t_in_v} \langle \, {\bf v} \, \gamma''' \, |t |
 \, {\bf v}'' \, \gamma'' \,
\rangle &=& \delta^{3}({\bf v}_{2}-{\bf v}''_{2})\, \delta^{3}({\bf
v}_{3}-{\bf v}''_{3})\,  \delta_{m''_{s_{3}} m'''_{s_{3}}} \,
\delta_{m''_{s_{4}} m'''_{s_{4}}} \, \delta_{m''_{t_{3}}
m'''_{t_{3}}} \, \delta_{m''_{t_{4}} m'''_{t_{4}}} \nonumber
\\ &\times&  \langle \,
{\bf v}_{1} \, m'''_{s_{1}} m'''_{s_{2}} \, m'''_{t_{1}}
m'''_{t_{2}} |t(\epsilon^{*}) |{\bf v}''_{1} \, m''_{s_{1}}
m''_{s_{2}} \, m''_{t_{1}} m''_{t_{2}} \, \rangle, \nonumber
\\  \epsilon^{*}&=&E-\frac{v_{2}^{2}}{2m}-\frac{v_{3}^{2}}{m},
\end{eqnarray}
\begin{eqnarray}
\label{Pbar_in_v} \langle  \, {\bf v}'' \, \gamma'' \, | \tilde{P}|
 \, {\bf v}' \, \gamma' \, \rangle &=&
\delta^{3}({\bf v}''_{1}-{\bf v}'_{3}) \, \delta^{3}({\bf
v}''_{2}+{\bf v}'_{2}) \, \delta^{3}({\bf v}''_{3} -{\bf v}'_{1})
\nonumber
\\ &\times&  \,
\delta_{m''_{s_{1}} m'_{s_{3}}} \delta_{m''_{s_{2}} m'_{s_{4}}}
\delta_{m''_{s_{3}} m'_{s_{1}}} \delta_{m''_{s_{4}} m'_{s_{2}}} \,
\nonumber
\\ &\times&  \, \delta_{m''_{t_{1}} m'_{t_{3}}} \delta_{m''_{t_{2}} m'_{t_{4}}}
\delta_{m''_{t_{3}} m'_{t_{1}}} \delta_{m''_{t_{4}} m'_{t_{2}}} .
\end{eqnarray}

Inserting Eqs. (\ref{t_in_v}) and (\ref{Pbar_in_v}) into Eq.
(\ref{G0_T_Pbar}) leads to:
\begin{eqnarray}
\langle \, {\bf v}\, \beta \, |G_{0}t \tilde{P} | \, {\bf v}' \,
\gamma' \, \rangle &=& \frac{\delta^{3}({\bf v}_{2}+{\bf
v}'_{2})\, \delta^{3}({\bf v}_{3}-{\bf v}'_{1})
}{E-\frac{v_{1}^{2}}{m}
-\frac{v_{2}^{2}}{2m}-\frac{v_{3}^{2}}{m}}\nonumber \\ &\times&
\sum_{\gamma'''} \, g_{\beta \gamma'''} \, \delta_{m'''_{s_{3}}
m'_{s_{1}}} \delta_{m'''_{s_{4}} m'_{s_{2}}} \,
\delta_{m'''_{t_{3}} m'_{t_{1}}} \delta_{m'''_{t_{4}} m'_{t_{2}}}
\nonumber \\ &\times& \langle \, {\bf v}_{1} \, m'''_{s_{1}}
m'''_{s_{2}} \, m'''_{t_{1}} m'''_{t_{2}} |t(\epsilon^{*}) |{\bf
v}'_{3} \, m'_{s_{3}} m'_{s_{4}} \, m'_{t_{3}} m'_{t_{4}} \,
\rangle.  \label{G0_T_Pbar_in_v}
\end{eqnarray}

For evaluation of the third term, Eq. (\ref{matrix_elements_3}), we
should use the relation between the Jacobi momenta in different
chains $(123,4)$ and $(124,3)$, which results:
\begin{eqnarray}
\label{1-P34_uu} \langle \, {\bf u}' \, \gamma' \, | 1-P_{34}| \,
{\bf u}'' \, \alpha'' \, \rangle &=& \, \sum_{\gamma''} \,
g_{\gamma'' \alpha''} \,
 \delta^{3}({\bf u}'_{1}-{\bf u}''_{1}) \, \nonumber \\ &\times&
 \Biggl\{ \delta^{3}({\bf u}'_{2}-{\bf u}''_{2})\,\delta^{3}({\bf
u}'_{3}-{\bf u}''_{3}) \, \delta_{\gamma'_{1234} \gamma''_{1234}}
\nonumber \\ && - \delta^{3}({\bf u}'_{2} -\frac{1}{3}{\bf
u}''_{2} -\frac{8}{9}{\bf u}''_{3} )\, \delta^{3}({\bf u}'_{3}
-{\bf u}''_{2} +\frac{1}{3}{\bf u}''_{3} ) \,
\delta_{\gamma'_{1234} \gamma''_{1243}}  \Biggr \}. \nonumber \\
\end{eqnarray}

And finally for the evaluation of the fourth and fifth terms, Eqs.
(\ref{matrix_elements_4}) and (\ref{matrix_elements_5}), we should
use the relation between the Jacobi momenta in two naturally
different chains $(123,4)$ and $(12,34)$. the result is:
\begin{eqnarray}
\label{uv} \langle \, {\bf u}' \, \gamma' \, |  \, {\bf v}' \,
\beta' \, \rangle = \,\, g_{\gamma' \beta'} \, \delta^{3}({\bf
v}'_{1}-{\bf u}'_{1}) \, \delta^{3}({\bf v}'_{2}+{\bf u}'_{2}+
\frac{2}{3}{\bf u}'_{3} )\, \delta^{3}({\bf v}'_{3}-\frac{1}{2}{\bf
u}'_{2}+\frac{2}{3}{\bf u}'_{3}),
\end{eqnarray}
\begin{eqnarray}
\label{1-P34_vu} \langle \, {\bf v}' \, \gamma' \, | 1-P_{34}| \,
{\bf u}' \, \alpha' \, \rangle &=& \, \sum_{\gamma''} \, g_{\gamma''
\alpha'} \,   \delta^{3}({\bf u}'_{1}-{\bf v}'_{1}) \nonumber \\
&\times&
  \Biggl\{\, \delta^{3}({\bf
u}'_{2}+\frac{2}{3}{\bf v}'_{2}- \frac{2}{3}{\bf v}'_{3} )\,\
\delta^{3}({\bf u}'_{3}+\frac{1}{2}{\bf v}'_{2}+{\bf v}'_{3}) \,
\delta_{\gamma'_{1234} \gamma''_{1234}} \,
 \nonumber \\ &&   \,+  \delta^{3}({\bf u}'_{2}+\frac{2}{3}{\bf v}'_{2}+
\frac{2}{3}{\bf v}'_{3} )\,\ \delta^{3}({\bf u}'_{3}+\frac{1}{2}{\bf
v}'_{2}-{\bf v}'_{3})
 \, \delta_{\gamma'_{1234} \gamma''_{1243}}   \Biggr\}. \nonumber \\
\end{eqnarray}

By these considerations, in the following we evaluate the first and
second Yakubovsky components separately. For the first component we
can rewrite Eq. (\ref{YCs_basis2}) by considering Eqs.
(\ref{1-P34_uu}) and (\ref{uv}) as:
\begin{eqnarray}
\label{YC1_basis2-p34}
 \langle \, {\bf u}\, \alpha \, |\psi_{1}\rangle &=&
  \intsum_{\gamma'}^{\bf u'} \, \intsum_{\alpha''}^{\bf u''} \, \langle \, {\bf u}\, \alpha \, |G_{0}t P |  \, {\bf u}'
\, \gamma' \, \rangle \, \sum_{\gamma''} \, g_{\gamma'' \alpha''} \,
 \delta^{3}({\bf u}'_{1}-{\bf u}''_{1}) \, \nonumber \\ && \times
  \Biggl\{\, \delta^{3}({\bf u}'_{2}-{\bf u}''_{2})\,\delta^{3}({\bf
u}'_{3}-{\bf u}''_{3}) \, \delta_{\gamma'_{1234} \gamma''_{1243}}
 \nonumber \\ &&  \quad - \delta^{3}({\bf u}'_{2} -\frac{1}{3}{\bf u}''_{2}
-\frac{8}{9}{\bf u}''_{3} )\, \delta^{3}({\bf u}'_{3} -{\bf
u}''_{2} +\frac{1}{3}{\bf u}''_{3} ) \, \delta_{\gamma'_{1234}
\gamma''_{1243}}  \, \Biggr \}  \nonumber \\ &&\times \langle \,
{\bf u}'' \, \alpha'' \, |\psi_{1}\rangle \nonumber
\\  &+&      \intsum_{\gamma'}^{\bf u'} \, \intsum_{\beta'}^{\bf v'}
 \,\ \langle \, {\bf u}\, \alpha \, |G_{0}t P | \, {\bf u}'
\, \gamma' \, \rangle \nonumber
\\ &&   \times
\,\, g_{\gamma' \beta'} \, \delta^{3}({\bf v}'_{1}-{\bf u}'_{1})
\, \delta^{3}({\bf v}'_{2}+{\bf u}'_{2}+ \frac{2}{3}{\bf u}'_{3}
)\, \delta^{3}({\bf v}'_{3}-\frac{1}{2}{\bf
u}'_{2}+\frac{2}{3}{\bf u}'_{3}) \nonumber \\ &&\times
 \langle \, {\bf v}' \, \beta' \,
|\psi_{2}\rangle,
\end{eqnarray}
by integrating over ${\bf u}''$ and ${\bf v}'$ vector sets in the
first and the second terms respectively, and considering Eq.
(\ref{G0_T_P_in_u}), we obtain:
\begin{eqnarray}
\label{YC1_basis2-G0tP}
 \langle \, {\bf u}\, \alpha \, |\psi_{1}\rangle &=&
  \intsum_{\gamma'}^{\bf u'} \, \frac{\delta^{3}({\bf u}_{3}-{\bf
u}'_{3})}{E-\frac{u_{1}^{2}}{m}
-\frac{3u_{2}^{2}}{4m}-\frac{2u_{3}^{2}}{3m}} \, \sum_{\gamma'''}
\, g_{\alpha \gamma'''} \, \delta_{m'''_{s_{4}} m'_{s_{4}}} \,
\delta_{m'''_{t_{4}} m'_{t_{4}}}  \nonumber \\ &\times&  \Biggl [
\, \delta^{3}({\bf u}_{2} -{\bf u}'_{1}+ \frac{1}{2}{\bf u}'_{2}
)\, \delta_{m'''_{s_{3}} m'_{s_{1}}} \, \delta_{m'''_{t_{3}}
m'_{t_{1}}} \, \nonumber \\ && \quad \times \langle {\bf u}_{1} \,
m'''_{s_{1}} m'''_{s_{2}} \, m'''_{t_{1}} m'''_{t_{2}}
|t(\epsilon) |\frac{-1}{2}{\bf u}_{2} -{\bf u}'_{2} \, m'_{s_{2}}
m'_{s_{3}} \, m'_{t_{2}} m'_{t_{3}} \rangle \nonumber \\ && \,\, +
\delta^{3}({\bf u}_{2} +{\bf u}'_{1}+ \frac{1}{2}{\bf u}'_{2} )\,
\delta_{m'''_{s_{3}} m'_{s_{2}}} \, \delta_{m'''_{t_{3}}
m'_{t_{2}}} \, \nonumber \\ && \quad \times \langle{\bf u}_{1} \,
m'''_{s_{1}} m'''_{s_{2}} \, m'''_{t_{1}} m'''_{t_{2}}
|t(\epsilon) |\frac{1}{2}{\bf u}_{2} +{\bf u}'_{2} \, m'_{s_{3}}
m'_{s_{1}} \, m'_{t_{3}} m'_{t_{1}} \rangle \, \Biggr ] \nonumber
\\ &\times& \Biggl\{\,\, \sum_{\alpha''} \,
g_{\gamma' \alpha''} \, \, \langle \, {\bf u}' \, \alpha'' \,
|\psi_{1}\rangle \nonumber \\ && \quad- \sum_{\alpha'',\gamma''}
\, g_{\gamma'' \alpha''} \, \delta
_{\gamma'_{1234}\gamma''_{1243}} \, \langle \, {\bf u}'_{1} \,\,
\frac{1}{3}{\bf u}'_{2}+\frac{8}{9}{\bf u}'_{3} \,\, {\bf
u}'_{2}-\frac{1}{3}{\bf u}'_{3} \,\, \alpha'' \, |\psi_{1}\rangle
\nonumber
\\ && \quad + \sum_{\beta'}  \,\, g_{\gamma' \beta'} \, \langle \, {\bf u}'_{1} \,\,
-{\bf u}'_{2}- \frac{2}{3}{\bf u}'_{3} \,\, +\frac{1}{2}{\bf
u}'_{2}-\frac{2}{3}{\bf u}'_{3} \, \beta' \, |\psi_{2}\rangle \,
\Biggr \},
\end{eqnarray}

and by integrating over ${\bf u}'_{1}$ and ${\bf u}'_{3}$ vectors
\begin{eqnarray}
\label{YC1_basis2-u'integration}
 \langle \, {\bf u}\, \alpha \, |\psi_{1}\rangle &=&
 \int d^{3}u'_{2} \, \sum_{\gamma',\gamma'''} \,
g_{\alpha \gamma'''} \,  \frac{\delta_{m'''_{s_{4}} m'_{s_{4}}} \,
\delta_{m'''_{t_{4}} m'_{t_{4}}}}{E-\frac{u_{1}^{2}}{m}
-\frac{3u_{2}^{2}}{4m}-\frac{2u_{3}^{2}}{3m}} \, \nonumber \\ &&
\hspace{-20mm}  \times
  \Biggl [ \,  \delta_{m'''_{s_{3}} m'_{s_{1}}}
\, \delta_{m'''_{t_{3}} m'_{t_{1}}} \, \langle {\bf u}_{1} \,
m'''_{s_{1}} m'''_{s_{2}} \, m'''_{t_{1}} m'''_{t_{2}}
|t(\epsilon) |\frac{-1}{2}{\bf u}_{2} -{\bf u}'_{2} \, m'_{s_{2}}
m'_{s_{3}} \, m'_{t_{2}} m'_{t_{3}} \rangle \nonumber \\ &&
\hspace{-20mm} \quad \times\Biggl\{\,\, \sum_{\alpha''} \,
g_{\gamma' \alpha''} \, \, \langle \, {\bf u}_{2}+\frac{1}{2}{\bf
u}'_{2} \,\, {\bf u}'_{2} \,\, {\bf u}_{3} \,\, \alpha'' \,
|\psi_{1}\rangle \nonumber \\ && \hspace{-20mm} \quad \quad-
\sum_{\alpha'',\gamma''}  \, g_{\gamma'' \alpha''} \, \delta
_{\gamma'_{1234}\gamma''_{1243}} \, \langle \, {\bf
u}_{2}+\frac{1}{2}{\bf u}'_{2} \,\, \frac{1}{3}{\bf
u}'_{2}+\frac{8}{9}{\bf u}_{3} \,\, {\bf u}'_{2}-\frac{1}{3}{\bf
u}_{3} \,\, \alpha'' \, |\psi_{1}\rangle \nonumber \\ &&
\hspace{-20mm} \quad \quad + \sum_{\beta'}  \,\, g_{\gamma'
\beta'} \, \langle \, {\bf u}_{2}+\frac{1}{2}{\bf u}'_{2} \,\,
-{\bf u}'_{2}- \frac{2}{3}{\bf u}_{3} \,\, +\frac{1}{2}{\bf
u}'_{2}-\frac{2}{3}{\bf u}_{3} \, \beta' \, |\psi_{2}\rangle \,
\Biggr \} \nonumber \\ && \hspace{-20mm}  \,\,+ \,
\delta_{m'''_{s_{3}} m'_{s_{2}}} \, \delta_{m'''_{t_{3}}
m'_{t_{2}}} \, \langle{\bf u}_{1} \, m'''_{s_{1}} m'''_{s_{2}} \,
m'''_{t_{1}} m'''_{t_{2}} |t(\epsilon) |\frac{1}{2}{\bf u}_{2}
+{\bf u}'_{2} \, m'_{s_{3}} m'_{s_{1}} \, m'_{t_{3}} m'_{t_{1}}
\rangle \nonumber \\ && \hspace{-20mm} \quad \times\Biggl\{\,\,
\sum_{\alpha''} \, g_{\gamma' \alpha''} \, \, \langle \, -{\bf
u}_{2}-\frac{1}{2}{\bf u}'_{2} \,\, {\bf u}'_{2} \,\, {\bf u}_{3}
\,\, \alpha'' \, |\psi_{1}\rangle \nonumber \\ && \hspace{-20mm}
\quad \quad- \sum_{\alpha'',\gamma''}  \, g_{\gamma'' \alpha''} \,
\delta _{\gamma'_{1234}\gamma''_{1243}} \, \langle \, -{\bf
u}_{2}-\frac{1}{2}{\bf u}'_{2} \,\, \frac{1}{3}{\bf
u}'_{2}+\frac{8}{9}{\bf u}_{3} \,\, {\bf u}'_{2}-\frac{1}{3}{\bf
u}_{3} \,\, \alpha'' \, |\psi_{1}\rangle \nonumber \\ &&
\hspace{-20mm} \quad\quad + \sum_{\beta'}  \,\, g_{\gamma' \beta'}
\, \langle \, -{\bf u}_{2}-\frac{1}{2}{\bf u}'_{2} \,\, -{\bf
u}'_{2}- \frac{2}{3}{\bf u}_{3} \,\, +\frac{1}{2}{\bf
u}'_{2}-\frac{2}{3}{\bf u}_{3} \, \beta' \, |\psi_{2}\rangle \,
\Biggr \} \, \Biggr ].
\end{eqnarray}

In order to evaluate the Eq. (\ref{YC1_basis2-u'integration}) we
consider the result of the permutation operators $P_{12}$ and
$P_{34}$ action on the Yakubovsky components, the space and also the
spin-isospin parts of the basis states as:
\begin{eqnarray}
\label{eq.p12-p34_application} P_{12} |\psi_{1}\rangle &=&
-|\psi_{1}\rangle, \nonumber \\ P_{12} |\psi_{2}\rangle &=&
-|\psi_{2}\rangle, \nonumber \\ P_{34} |\psi_{2}\rangle &=&
-|\psi_{2}\rangle, \nonumber \\ P_{12} | \, {\bf u}\, \rangle &=&
| \, -{\bf u}_{1} \, {\bf u}_{2}\, {\bf u}_{3}\,\rangle, \nonumber
\\ P_{12} | \, {\bf v}\, \rangle &=& | \, -{\bf v}_{1} \, {\bf
v}_{2}\, {\bf v}_{3}\,\rangle,  \nonumber
\\ P_{34} | \, {\bf v}\, \rangle &=& | \, {\bf v}_{1} \, {\bf
v}_{2}\, -{\bf v}_{3}\,\rangle, \nonumber
\\P_{12} | \, \alpha \rangle &=&
(-)^{s_{1}+s_{2}-s_{12}}(-)^{t_{1}+t_{2}-t_{12}} | \, \alpha \rangle
= (-)^{s_{12}+t_{12}} |\, \alpha \rangle,  \nonumber
\\P_{12} | \, \beta \rangle &=&
(-)^{s_{1}+s_{2}-s_{12}}(-)^{t_{1}+t_{2}-t_{12}} | \, \beta \rangle
 = (-)^{s_{12}+t_{12}} |\, \beta \rangle, \nonumber
\\P_{34} | \, \beta \rangle &=&
(-)^{s_{3}+s_{4}-s_{34}}(-)^{t_{3}+t_{4}-t_{34}} | \, \beta \rangle
= (-)^{s_{34}+t_{34}} |\, \beta \rangle, \nonumber
\\ P_{12} | \,  \gamma \rangle  &=&  | \, m_{s_{2}} m_{s_{1}} m_{s_{3}} m_{s_{4}} \,
m_{t_{2}} m_{t_{1}} m_{t_{3}} m_{t_{4}} \rangle \equiv | \,
\gamma_{2134} \rangle, \nonumber
\\ P_{34} | \,  \gamma \rangle  &=&  | \, m_{s_{1}} m_{s_{2}} m_{s_{4}} m_{s_{3}} \,
m_{t_{1}} m_{t_{2}} m_{t_{4}} m_{t_{3}} \rangle \equiv | \,
\gamma_{1243} \rangle.
\end{eqnarray}

 By taking the Eq. (\ref{eq.p12-p34_application}) into consideration
the matrix elements of the NN $t$-matrix as well as the Yakubovsky
components in the second term can be rewritten as:
\begin{eqnarray}
\label{t-in-u-revised} && \langle{\bf u}_{1} \, m'''_{s_{1}}
m'''_{s_{2}} \, m'''_{t_{1}} m'''_{t_{2}} |t(\epsilon)
|\frac{1}{2}{\bf u}_{2} +{\bf u}'_{2} \, m'_{s_{3}} m'_{s_{1}} \,
m'_{t_{3}} m'_{t_{1}} \rangle \nonumber \\ && \equiv  \langle{\bf
u}_{1} \, m'''_{s_{1}} m'''_{s_{2}} \, m'''_{t_{1}} m'''_{t_{2}}
|t(\epsilon) P_{12} P_{12}|\frac{1}{2}{\bf u}_{2} +{\bf u}'_{2} \,
m'_{s_{3}} m'_{s_{1}} \, m'_{t_{3}} m'_{t_{1}} \rangle \nonumber
\\ && =  \langle{\bf u}_{1} \, m'''_{s_{1}} m'''_{s_{2}} \,
m'''_{t_{1}} m'''_{t_{2}} |t(\epsilon)  P_{12}|-\frac{1}{2}{\bf
u}_{2} -{\bf u}'_{2} \, m'_{s_{1}} m'_{s_{3}} \, m'_{t_{1}}
m'_{t_{3}} \rangle,
\end{eqnarray}
\begin{eqnarray}
\label{YC1_revised_1} && \sum_{\alpha''} \, g_{\gamma' \alpha''} \,
\, \langle \, -{\bf u}_{2}-\frac{1}{2}{\bf u}'_{2} \,\, {\bf u}'_{2}
\,\, {\bf u}_{3} \,\, \alpha'' \, |\psi_{1}\rangle \nonumber
\\ && \equiv
\sum_{\alpha''} \, g_{\gamma' \alpha''} \, \, \langle \, -{\bf
u}_{2}-\frac{1}{2}{\bf u}'_{2} \,\, {\bf u}'_{2} \,\, {\bf u}_{3}
\,\, \alpha'' \, |P_{12} P_{12}|\psi_{1}\rangle \nonumber
\\    && = - (-)^{s_{12}''+t_{12}''}
\sum_{\alpha''} \, g_{\gamma' \alpha''} \, \, \langle \, {\bf
u}_{2}+\frac{1}{2}{\bf u}'_{2} \,\, {\bf u}'_{2} \,\, {\bf u}_{3}
\,\, \alpha'' \, |\psi_{1}\rangle,
\end{eqnarray}
\begin{eqnarray}
\label{YC1_revised_2} &&  \sum_{\alpha'',\gamma''}  \, g_{\gamma''
\alpha''} \, \delta _{\gamma'_{1234}\gamma''_{1243}} \, \langle \,
-{\bf u}_{2}-\frac{1}{2}{\bf u}'_{2} \,\, \frac{1}{3}{\bf
u}'_{2}+\frac{8}{9}{\bf u}_{3} \,\, {\bf u}'_{2}-\frac{1}{3}{\bf
u}_{3} \,\, \alpha'' \, |\psi_{1}\rangle  \nonumber
\\   && \equiv \sum_{\alpha'',\gamma''}
\, g_{\gamma'' \alpha''} \, \delta _{\gamma'_{1234}\gamma''_{1243}}
\, \langle \, -{\bf u}_{2}-\frac{1}{2}{\bf u}'_{2} \,\,
\frac{1}{3}{\bf u}'_{2}+\frac{8}{9}{\bf u}_{3} \,\, {\bf
u}'_{2}-\frac{1}{3}{\bf u}_{3} \,\, \alpha'' \, |
P_{12}P_{12}|\psi_{1}\rangle \nonumber
\\   && = - (-)^{s_{12}''+t_{12}''} \sum_{\alpha'',\gamma''}
 \, g_{\gamma'' \alpha''} \, \delta
_{\gamma'_{1234}\gamma''_{1243}} \nonumber
\\   && \quad \times \langle \, {\bf
u}_{2}+\frac{1}{2}{\bf u}'_{2} \,\, \frac{1}{3}{\bf
u}'_{2}+\frac{8}{9}{\bf u}_{3} \,\, {\bf u}'_{2}-\frac{1}{3}{\bf
u}_{3} \,\, \alpha'' \, | P_{12}P_{12}|\psi_{1}\rangle,
\end{eqnarray}
\begin{eqnarray}
\label{YC1_revised_3}
  && \sum_{\beta'}  \,\, g_{\gamma' \beta'} \,
\langle \, -{\bf u}_{2}-\frac{1}{2}{\bf u}'_{2} \,\, -{\bf u}'_{2}-
\frac{2}{3}{\bf u}_{3} \,\, +\frac{1}{2}{\bf u}'_{2}-\frac{2}{3}{\bf
u}_{3} \, \beta' \, |\psi_{2}\rangle \nonumber
\\  && \equiv \sum_{\beta'}  \,\, g_{\gamma' \beta'} \, \langle \, -{\bf u}_{2}-\frac{1}{2}{\bf u}'_{2} \,\,
-{\bf u}'_{2}- \frac{2}{3}{\bf u}_{3} \,\, +\frac{1}{2}{\bf
u}'_{2}-\frac{2}{3}{\bf u}_{3} \, \beta' \, | P_{12} P_{12}
|\psi_{2}\rangle \nonumber
\\  && = - (-)^{s_{12}'+t_{12}'} \sum_{\beta'}  \,\, g_{\gamma' \beta'} \, \langle \, {\bf u}_{2}+\frac{1}{2}{\bf u}'_{2} \,\,
-{\bf u}'_{2}- \frac{2}{3}{\bf u}_{3} \,\, +\frac{1}{2}{\bf
u}'_{2}-\frac{2}{3}{\bf u}_{3} \, \beta' \, | P_{12} P_{12}
|\psi_{2}\rangle. \nonumber
\\
\end{eqnarray}

After inserting Eqs. (\ref{t-in-u-revised})-(\ref{YC1_revised_3}) in
Eq. (\ref{YC1_basis2-u'integration}) we obtain:
\begin{eqnarray}
\label{YC1_revised2}
 \langle \, {\bf u}\, \alpha \, |\psi_{1}\rangle &=&
   \int d^{3}u'_{2} \, \sum_{\gamma',\gamma'''} \,
g_{\alpha \gamma'''} \,  \frac{\delta_{m'''_{s_{4}} m'_{s_{4}}} \,
\delta_{m'''_{t_{4}} m'_{t_{4}}}}{E-\frac{u_{1}^{2}}{m}
-\frac{3u_{2}^{2}}{4m}-\frac{2u_{3}^{2}}{3m}} \, \nonumber \\ &&
\hspace{-30mm} \times
  \Biggl [  \, \langle {\bf
u}_{1} \, m'''_{s_{1}} m'''_{s_{2}} \, m'''_{t_{1}} m'''_{t_{2}}
|t(\epsilon) |\frac{-1}{2}{\bf u}_{2} -{\bf u}'_{2} \, m'_{s_{2}}
m'_{s_{3}} \, m'_{t_{2}} m'_{t_{3}} \rangle \,\,
\delta_{m'''_{s_{3}} m'_{s_{1}}} \, \delta_{m'''_{t_{3}}
m'_{t_{1}}} \nonumber \\ && \hspace{-30mm} \quad
\times\Biggl\{\,\, \sum_{\alpha''} \, g_{\gamma' \alpha''} \, \,
\langle \, {\bf u}_{2}+\frac{1}{2}{\bf u}'_{2} \,\, {\bf u}'_{2}
\,\, {\bf u}_{3} \,\, \alpha'' \, |\psi_{1}\rangle \nonumber
\\ && \hspace{-30mm} \quad \quad  - \sum_{\alpha'', \gamma''}  \,
g_{\gamma'' \alpha''} \, \delta _{\gamma'_{1234}\gamma''_{1243}} \,
\langle \, {\bf u}_{2}+\frac{1}{2}{\bf u}'_{2} \,\, \frac{1}{3}{\bf
u}'_{2}+\frac{8}{9}{\bf u}_{3} \,\, {\bf u}'_{2}-\frac{1}{3}{\bf
u}_{3} \,\, \alpha'' \, |\psi_{1}\rangle \nonumber
\\ && \hspace{-30mm} \quad \quad + \sum_{\beta'}  \,\, g_{\gamma' \beta'} \, \langle \, {\bf u}_{2}+\frac{1}{2}{\bf u}'_{2} \,\,
-{\bf u}'_{2}- \frac{2}{3}{\bf u}_{3} \,\, +\frac{1}{2}{\bf
u}'_{2}-\frac{2}{3}{\bf u}_{3} \, \beta' \, |\psi_{2}\rangle \,
\Biggr \} \nonumber \\ && \hspace{-30mm}  \,\,+  \, \langle{\bf
u}_{1} \, m'''_{s_{1}} m'''_{s_{2}} \, m'''_{t_{1}} m'''_{t_{2}}
|t(\epsilon) P_{12}|-\frac{1}{2}{\bf u}_{2} -{\bf u}'_{2} \,
m'_{s_{1}} m'_{s_{3}} \, m'_{t_{1}} m'_{t_{3}} \rangle \,\,
\delta_{m'''_{s_{3}} m'_{s_{2}}} \, \delta_{m'''_{t_{3}}
m'_{t_{2}}} \nonumber \\ && \hspace{-30mm} \quad
\times\Biggl\{\,\, \sum_{\alpha''} \, g_{\gamma' \alpha''}  \,
\left(-(-)^{s_{12}''+t_{12}''}\right) \langle \, {\bf
u}_{2}+\frac{1}{2}{\bf u}'_{2} \,\, {\bf u}'_{2} \,\, {\bf u}_{3}
\,\, \alpha'' \, |\psi_{1}\rangle \nonumber
\\ && \hspace{-30mm} \quad \quad- \sum_{\alpha'', \gamma''} \,
g_{\gamma'' \alpha''} \, \delta _{\gamma'_{1234}\gamma''_{1243}} \,
\left(-(-)^{s_{12}''+t_{12}''}\right) \langle \, {\bf
u}_{2}+\frac{1}{2}{\bf u}'_{2} \,\, \frac{1}{3}{\bf
u}'_{2}+\frac{8}{9}{\bf u}_{3} \,\, {\bf u}'_{2}-\frac{1}{3}{\bf
u}_{3} \,\, \alpha'' \, |\psi_{1}\rangle \nonumber
\\ && \hspace{-30mm} \quad \quad + \sum_{\beta'}  \, g_{\gamma' \beta'}
 \, \left(-(-)^{s_{12}'+t_{12}'}\right) \langle \, {\bf u}_{2}+\frac{1}{2}{\bf u}'_{2} \,\,
-{\bf u}'_{2}- \frac{2}{3}{\bf u}_{3} \,\, +\frac{1}{2}{\bf
u}'_{2}-\frac{2}{3}{\bf u}_{3} \, \beta' \, |\psi_{2}\rangle \,
\Biggr \} \, \Biggr ], \nonumber
\\
\end{eqnarray}

The exchange of the labels $m'_{s_{1}}, m'_{t_{1}}$ to $m'_{s_{2}},
m'_{t_{2}}$ and reverse of it in the second term and consequently
the following relations:
\begin{eqnarray}
\label{CG_exchanged-labels} g_{\gamma' \alpha''} &\rightarrow&
(-)^{s_{12}''+t_{12}''} \, g_{\gamma' \alpha''} \nonumber \\
g_{\gamma'_{1243} \alpha''} &\rightarrow& (-)^{s_{12}''+t_{12}''}
\, g_{\gamma'_{1243} \alpha''} \nonumber \\ g_{\gamma' \beta'}
&\rightarrow& (-)^{s_{12}'+t_{12}'} \, g_{\gamma' \beta'}
\end{eqnarray}
lead to:
\begin{eqnarray}
\label{YC1-final}
 \langle \, {\bf u}\, \alpha \, |\psi_{1}\rangle &=&
   \int d^{3}u'_{2} \, \sum_{\gamma',\gamma'''} \,
g_{\alpha \gamma'''} \,  \frac{\delta_{m'''_{s_{4}} m'_{s_{4}}} \,
\delta_{m'''_{t_{4}} m'_{t_{4}}}}{E-\frac{u_{1}^{2}}{m}
-\frac{3u_{2}^{2}}{4m}-\frac{2u_{3}^{2}}{3m}} \, \,
\delta_{m'''_{s_{3}} m'_{s_{1}}} \, \delta_{m'''_{t_{3}}
m'_{t_{1}}} \nonumber \\ && \hspace{-15mm}\times     \, \langle
{\bf u}_{1} \, m'''_{s_{1}} m'''_{s_{2}} \, m'''_{t_{1}}
m'''_{t_{2}} |t(\epsilon)\, (1-P_{12}) |\frac{-1}{2}{\bf u}_{2}
-{\bf u}'_{2} \, m'_{s_{2}} m'_{s_{3}} \, m'_{t_{2}} m'_{t_{3}}
\rangle \nonumber \\ && \hspace{-15mm}\times \Biggl\{\,\,
\sum_{\alpha''} \, g_{\gamma' \alpha''} \, \, \langle \, {\bf
u}_{2}+\frac{1}{2}{\bf u}'_{2} \,\, {\bf u}'_{2} \,\, {\bf u}_{3}
\,\, \alpha'' \, |\psi_{1}\rangle \nonumber \\ && \hspace{-15mm}
\,\,\,- \sum_{\alpha'',\gamma''} \,
 g_{\gamma'' \alpha''} \, \delta
_{\gamma'_{1234}\gamma''_{1243}} \, \langle \, {\bf
u}_{2}+\frac{1}{2}{\bf u}'_{2} \,\, \frac{1}{3}{\bf
u}'_{2}+\frac{8}{9}{\bf u}_{3} \,\, {\bf u}'_{2}-\frac{1}{3}{\bf
u}_{3} \,\, \alpha'' \, |\psi_{1}\rangle \nonumber \\ &&
\hspace{-15mm} \,\,   \,\, + \sum_{\beta'}  \,\, g_{\gamma'
\beta'} \, \langle \, {\bf u}_{2}+\frac{1}{2}{\bf u}'_{2} \,\,
-{\bf u}'_{2}- \frac{2}{3}{\bf u}_{3} \,\, +\frac{1}{2}{\bf
u}'_{2}-\frac{2}{3}{\bf u}_{3} \, \beta' \, |\psi_{2}\rangle \,
\Biggr \},
\end{eqnarray}

For the second component we can rewrite Eq. (\ref{YCs_basis2}) by
considering Eq. (\ref{1-P34_vu}) as:
\begin{eqnarray}
\label{YC2-p34}
 \langle \, {\bf v}\, \beta \, |\psi_{2}\rangle
&=& \intsum_{\gamma'}^{\bf v'} \, \intsum_{\alpha'}^{\bf u'} \,
\langle \, {\bf v}\, \beta \, |G_{0}t \tilde{P}| \, {\bf v}' \,
\gamma' \, \rangle
  \, \sum_{\gamma''} \, g_{\gamma'' \alpha'} \,
\delta^{3}({\bf u}'_{1}-{\bf v}'_{1}) \nonumber \\ &&
\hspace{-15mm}\times
   \Biggl\{\,\, \delta^{3}({\bf
u}'_{2}+\frac{2}{3}{\bf v}'_{2}- \frac{2}{3}{\bf v}'_{3} )\,\
\delta^{3}({\bf u}'_{3}+\frac{1}{2}{\bf v}'_{2}+{\bf v}'_{3}) \,
\delta_{\gamma'_{1234} \gamma''_{1234}} \,
 \nonumber \\ && \hspace{-15mm}  \quad +  \delta^{3}({\bf u}'_{2}+\frac{2}{3}{\bf v}'_{2}+
\frac{2}{3}{\bf v}'_{3} )\,\ \delta^{3}({\bf u}'_{3}+\frac{1}{2}{\bf
v}'_{2}-{\bf v}'_{3})
 \, \delta_{\gamma'_{1234} \gamma''_{1243}}  \, \Biggr\}
   \times\langle \, {\bf u}' \, \alpha' \,
|\psi_{1}\rangle
  \nonumber \\ && \hspace{-15mm}+  \, \intsum_{\beta'}^{\bf v'} \, \langle \, {\bf v}\, \beta \, |G_{0}t \tilde{P}| \, {\bf
v}' \, \beta' \, \rangle \langle \, {\bf v}' \, \beta' \,
|\psi_{2}\rangle.
\end{eqnarray}

In this stage by integrating over ${\bf u}'$ vector set and
considering Eq. (\ref{G0_T_Pbar_in_v}) we obtain:
\begin{eqnarray}
\label{YC2-G0tpibar}
 \langle \, {\bf v}\, \beta \, |\psi_{2}\rangle
&=& \, \intsum_{\gamma'}^{\bf v'} \,  \frac{\delta^{3}({\bf
v}_{2}+{\bf v}'_{2})\, \delta^{3}({\bf v}_{3}-{\bf v}'_{1})
}{E-\frac{v_{1}^{2}}{m} -\frac{v_{2}^{2}}{2m}-\frac{v_{3}^{2}}{m}}
\,\, \nonumber \\ &\times&  \sum_{\gamma'''} \, g_{\beta
\gamma'''} \, \delta_{m'''_{s_{3}} m'_{s_{1}}}
\delta_{m'''_{s_{4}} m'_{s_{2}}} \, \delta_{m'''_{t_{3}}
m'_{t_{1}}} \delta_{m'''_{t_{4}} m'_{t_{2}}} \nonumber \\
&\times& \langle \, {\bf v}_{1} \, m'''_{s_{1}} m'''_{s_{2}} \,
m'''_{t_{1}} m'''_{t_{2}} |t(\epsilon^{*}) |{\bf v}'_{3} \,
m'_{s_{3}} m'_{s_{4}} \, m'_{t_{3}} m'_{t_{4}} \, \rangle
\nonumber \\ &\times&
  \Biggl\{\,\,  \sum_{\alpha'} \, g_{\gamma' \alpha'} \, \langle \, {\bf v}'_{1} \,\, -\frac{2}{3}{\bf
v}'_{2}+ \frac{2}{3}{\bf v}'_{3} \,\, -\frac{1}{2}{\bf v}'_{2}-{\bf
v}'_{3}  \,\, \alpha' \, |\psi_{1}\rangle
 \nonumber \\ &&   \, \,\,\, \,- \sum_{\alpha',\gamma''} \, g_{\gamma'' \alpha'} \, \delta_{\gamma'_{1234} \gamma''_{1243}}
  \, \langle \, {\bf v}'_{1} \,\, -\frac{2}{3}{\bf
v}'_{2}- \frac{2}{3}{\bf v}'_{3} \,\, -\frac{1}{2}{\bf v}'_{2}+{\bf
v}'_{3}  \,\, \alpha' \, |\psi_{1}\rangle
  \nonumber \\ &&  \, \,\,\, \,+ \sum_{\beta'} \, g_{\gamma' \beta'} \,  \langle \, {\bf v}' \, \beta' \,
|\psi_{2}\rangle \, \Biggr\},
\end{eqnarray}
and in the next stage by integrating over ${\bf v}'_{1}$ and ${\bf
v}'_{2}$ vectors we arrive to the following equation:
\begin{eqnarray}
\label{YC2-v'-integration}
 \langle \, {\bf v}\, \beta \, |\psi_{2}\rangle
&=&   \int d^{3}v'_{3} \, \sum_{\gamma',\gamma'''} \, g_{\beta
\gamma'''} \, \frac{\delta_{m'''_{s_{3}} m'_{s_{1}}}
\delta_{m'''_{s_{4}} m'_{s_{2}}} \, \delta_{m'''_{t_{3}} m'_{t_{1}}}
\delta_{m'''_{t_{4}} m'_{t_{2}}} }{E-\frac{v_{1}^{2}}{m}
-\frac{v_{2}^{2}}{2m}-\frac{v_{3}^{2}}{m}} \,\,  \nonumber \\
&\times& \langle \, {\bf v}_{1} \, m'''_{s_{1}} m'''_{s_{2}} \,
m'''_{t_{1}} m'''_{t_{2}} |t(\epsilon^{*}) |{\bf v}'_{3} \,
m'_{s_{3}} m'_{s_{4}} \, m'_{t_{3}} m'_{t_{4}} \, \rangle \nonumber
\\ &\times&
  \Biggl\{\,\,  \sum_{\alpha'} \, g_{\gamma' \alpha'} \, \langle \, {\bf v}_{3} \,\, \frac{2}{3}{\bf
v}_{2}+ \frac{2}{3}{\bf v}'_{3} \,\, \frac{1}{2}{\bf v}_{2}-{\bf
v}'_{3}  \,\, \alpha' \, |\psi_{1}\rangle
 \nonumber \\ &&   \, \,\,\, \,- \sum_{\alpha'} \, g_{\gamma'_{1243} \alpha'}
  \, \langle \, {\bf v}_{3} \,\, \frac{2}{3}{\bf
v}_{2}- \frac{2}{3}{\bf v}'_{3} \,\, \frac{1}{2}{\bf v}_{2}+{\bf
v}'_{3}  \,\, \alpha' \, |\psi_{1}\rangle
  \nonumber \\ &&  \, \,\,\, \,+ \sum_{\beta'} \, g_{\gamma' \beta'} \,  \langle \, {\bf v}_{3} \,\, -{\bf v}_{2} \,\, {\bf v}'_{3} \,\, \beta' \,
|\psi_{2}\rangle \, \Biggr\}.
\end{eqnarray}

Therefore by considering the result of the permutation operators
$P_{12}$ and $P_{34}$ action, Eq. (\ref{eq.p12-p34_application}),
the matrix elements of the NN $t$-matrix as well as the Yakubovsky
components, under the exchange of the labels $m'_{s_{3}},
m'_{t_{3}}$ to $m'_{s_{4}}, m'_{t_{4}}$ and reverse of it with
changing ${\bf v}'_{3}$ to $-{\bf v}'_{3}$, can be obtained as:
\begin{eqnarray}
\label{YC2-t-exchange}
 && \langle \, {\bf v}_{1} \, m'''_{s_{1}}
m'''_{s_{2}} \, m'''_{t_{1}} m'''_{t_{2}} |t(\epsilon^{*}) |{\bf
v}'_{3} \, m'_{s_{3}} m'_{s_{4}} \, m'_{t_{3}} m'_{t_{4}} \,
\rangle \nonumber \\ && \dashrightarrow\, \langle \, {\bf v}_{1}
\, m'''_{s_{1}} m'''_{s_{2}} \, m'''_{t_{1}} m'''_{t_{2}}
|t(\epsilon^{*}) |-{\bf v}'_{3} \, m'_{s_{4}} m'_{s_{3}} \,
m'_{t_{4}} m'_{t_{3}} \, \rangle \nonumber \\ && = \langle \, {\bf
v}_{1} \, m'''_{s_{1}} m'''_{s_{2}} \, m'''_{t_{1}} m'''_{t_{2}}
|t(\epsilon^{*}) P_{12} P_{12}|-{\bf v}'_{3} \, m'_{s_{4}}
m'_{s_{3}} \, m'_{t_{4}} m'_{t_{3}} \, \rangle \nonumber
\\ && = \langle \, {\bf v}_{1} \, m'''_{s_{1}} m'''_{s_{2}} \,
m'''_{t_{1}} m'''_{t_{2}} |t(\epsilon^{*}) P_{12} |{\bf v}'_{3} \,
m'_{s_{3}} m'_{s_{4}} \, m'_{t_{3}} m'_{t_{4}} \, \rangle
\end{eqnarray}
\begin{eqnarray}
\label{YC2-YCs-exchange1}
  && \sum_{\alpha'} \, g_{\gamma' \alpha'} \, \langle \, {\bf v}_{3} \,\, \frac{2}{3}{\bf
v}_{2}+ \frac{2}{3}{\bf v}'_{3} \,\, \frac{1}{2}{\bf v}_{2}-{\bf
v}'_{3}  \,\, \alpha' \, |\psi_{1}\rangle \nonumber \\ &&
 \dashrightarrow\, \sum_{\alpha'} \,
g_{\gamma'_{1243} \alpha'} \,  \langle \, {\bf v}_{3} \,\,
\frac{2}{3}{\bf v}_{2}- \frac{2}{3}{\bf v}'_{3} \,\, \frac{1}{2}{\bf
v}_{2}+{\bf v}'_{3}  \,\, \alpha' \, |\psi_{1}\rangle,
\end{eqnarray}
\begin{eqnarray}
\label{YC2-YCs-exchange2}
 &&    \sum_{\alpha'} \, g_{\gamma'_{1243} \alpha'}
  \, \langle \, {\bf v}_{3} \,\, \frac{2}{3}{\bf
v}_{2}- \frac{2}{3}{\bf v}'_{3} \,\, \frac{1}{2}{\bf v}_{2}+{\bf
v}'_{3}  \,\, \alpha' \, |\psi_{1}\rangle \nonumber \\ &&
 \dashrightarrow\, \sum_{\alpha'} \, g_{\gamma'
\alpha'}
  \, \langle \, {\bf v}_{3} \,\, \frac{2}{3}{\bf
v}_{2}+ \frac{2}{3}{\bf v}'_{3} \,\, \frac{1}{2}{\bf v}_{2}-{\bf
v}'_{3}  \,\, \alpha' \, |\psi_{1}\rangle,
\end{eqnarray}
\begin{eqnarray}
\label{YC2-YCs-exchange3}
  &&   \sum_{\beta'} \, g_{\gamma' \beta'} \,  \langle \, {\bf v}_{3} \,\, -{\bf v}_{2} \,\, {\bf v}'_{3} \,\, \beta' \,
|\psi_{2}\rangle \,\dashrightarrow\, \sum_{\beta'} \,
g_{\gamma'_{1243} \beta'} \,  \langle \, {\bf v}_{3} \,\, -{\bf
v}_{2} \,\, -{\bf v}'_{3} \,\, \beta' \, |\psi_{2}\rangle \nonumber
\\ &&  \equiv \sum_{\beta'} \, (-)^{s_{34}'+t_{34}'} \,
g_{\gamma' \beta'} \,  \langle \, {\bf v}_{3} \,\, -{\bf v}_{2} \,\,
-{\bf v}'_{3} \,\, \beta' \, |P_{34} P_{34}|\psi_{2}\rangle
\nonumber
\\ && \equiv \sum_{\beta'} \, (-)^{s_{34}'+t_{34}'} \,
g_{\gamma' \beta'} \,  \left( -(-)^{s_{34}'+t_{34}'}  \right) \,
\langle \, {\bf v}_{3} \,\, -{\bf v}_{2} \,\, {\bf v}'_{3} \,\,
\beta' \, |\psi_{2}\rangle \nonumber
\\ && \equiv - \sum_{\beta'} \,
g_{\gamma' \beta'} \,  \langle \, {\bf v}_{3} \,\, -{\bf v}_{2} \,\,
{\bf v}'_{3} \,\, \beta' \, |\psi_{2}\rangle.
\end{eqnarray}

With these considerations we can rewrite the Eq.
(\ref{YC2-v'-integration}) as:
\begin{eqnarray}
\label{YC2-revised}
 \langle \, {\bf v}\, \beta \, |\psi_{2}\rangle
&=&  \int d^{3}v'_{3} \, \sum_{\gamma',\gamma'''} \, g_{\beta
\gamma'''} \, \frac{\delta_{m'''_{s_{3}} m'_{s_{1}}}
\delta_{m'''_{s_{4}} m'_{s_{2}}} \, \delta_{m'''_{t_{3}} m'_{t_{1}}}
\delta_{m'''_{t_{4}} m'_{t_{2}}} }{E-\frac{v_{1}^{2}}{m}
-\frac{v_{2}^{2}}{2m}-\frac{v_{3}^{2}}{m}} \,\,  \nonumber \\
&\times& \langle \, {\bf v}_{1} \, m'''_{s_{1}} m'''_{s_{2}} \,
m'''_{t_{1}} m'''_{t_{2}} |t(\epsilon^{*}) P_{12}|{\bf v}'_{3} \,
m'_{s_{3}} m'_{s_{4}} \, m'_{t_{3}} m'_{t_{4}} \, \rangle \nonumber
\\ &\times& (-)
  \Biggl\{\,\,  \sum_{\alpha'} \, g_{\gamma' \alpha'} \, \langle \, {\bf v}_{3} \,\, \frac{2}{3}{\bf
v}_{2}+ \frac{2}{3}{\bf v}'_{3} \,\, \frac{1}{2}{\bf v}_{2}-{\bf
v}'_{3}  \,\, \alpha' \, |\psi_{1}\rangle
 \nonumber \\ &&   \,\,\,\, \,\,\,\,\,\,\,\,\,\, - \sum_{\alpha'} \, g_{\gamma'_{1234}  \alpha'}
  \, \langle \, {\bf v}_{3} \,\, \frac{2}{3}{\bf
v}_{2}- \frac{2}{3}{\bf v}'_{3} \,\, \frac{1}{2}{\bf v}_{2}+{\bf
v}'_{3}  \,\, \alpha' \, |\psi_{1}\rangle
  \nonumber \\ &&  \,\,\,\, \,\,\,\,\,\,\,\,\,\,+ \sum_{\beta'} \, g_{\gamma' \beta'} \,  \langle \, {\bf v}_{3} \,\, -{\bf v}_{2} \,\, {\bf v}'_{3} \,\, \beta' \,
|\psi_{2}\rangle \, \Biggr\},
\end{eqnarray}
and finally considering the Eqs. (\ref{YC2-v'-integration}) and
 (\ref{YC2-revised}) together leads to:
\begin{eqnarray}
\label{YC2-final}
 \langle \, {\bf v}\, \beta \, |\psi_{2}\rangle
&=& \frac{1}{2} \int d^{3}v'_{3} \, \sum_{\gamma',\gamma'''} \,
g_{\beta \gamma'''} \, \frac{\delta_{m'''_{s_{3}} m'_{s_{1}}}
\delta_{m'''_{s_{4}} m'_{s_{2}}} \, \delta_{m'''_{t_{3}} m'_{t_{1}}}
\delta_{m'''_{t_{4}} m'_{t_{2}}} }{E-\frac{v_{1}^{2}}{m}
-\frac{v_{2}^{2}}{2m}-\frac{v_{3}^{2}}{m}} \,\,  \nonumber \\
&\times& \langle \, {\bf v}_{1} \, m'''_{s_{1}} m'''_{s_{2}} \,
m'''_{t_{1}} m'''_{t_{2}} |t(\epsilon^{*}) (1-P_{12})|{\bf v}'_{3}
\, m'_{s_{3}} m'_{s_{4}} \, m'_{t_{3}} m'_{t_{4}} \, \rangle
\nonumber
\\ &\times&
  \Biggl\{\,\,  \sum_{\alpha'} \, g_{\gamma' \alpha'} \, \langle \, {\bf v}_{3} \,\, \frac{2}{3}{\bf
v}_{2}+ \frac{2}{3}{\bf v}'_{3} \,\, \frac{1}{2}{\bf v}_{2}-{\bf
v}'_{3}  \,\, \alpha' \, |\psi_{1}\rangle
 \nonumber \\ &&   \, \,\,\, \,- \sum_{\alpha'} \, g_{\gamma'_{1234}  \alpha'}
  \, \langle \, {\bf v}_{3} \,\, \frac{2}{3}{\bf
v}_{2}- \frac{2}{3}{\bf v}'_{3} \,\, \frac{1}{2}{\bf v}_{2}+{\bf
v}'_{3}  \,\, \alpha' \, |\psi_{1}\rangle
  \nonumber \\ &&  \, \,\,\, \,+ \sum_{\beta'} \, g_{\gamma' \beta'} \,  \langle \, {\bf v}_{3} \,\, -{\bf v}_{2} \,\, {\bf v}'_{3} \,\, \beta' \,
|\psi_{2}\rangle \, \Biggr\}.
\end{eqnarray}

By introducing the physical representation of the NN $t$-matrix (
see appendix (B) of Ref. \cite{Bayegan-PRC77}):
\begin{eqnarray} \label{eq.ta}
  _{a}\langle{\bf p} \, m_{s_{1}} m_{s_{2}} \, m_{t_{1}}
m_{t_{2}} |t(\varepsilon)| {\bf p}' \, m'_{s_{1}} m'_{s_{2}} \,
m'_{t_{1}} m'_{t_{2}} \rangle _{a} && \nonumber
\\ && \hspace{-60mm}= \langle{\bf p} \, m_{s_{1}} m_{s_{2}}
\, m_{t_{1}} m_{t_{2}} |t(\varepsilon)(1-P_{12})| {\bf p}' \,
m'_{s_{1}} m'_{s_{2}} \, m'_{t_{1}} m'_{t_{2}} \rangle,
\end{eqnarray}
the final representation of the three dimensional Yakubovsky
integral equations can be obtained by rewriting Eqs.
(\ref{YC1-final}) and (\ref{YC2-final}):
\begin{eqnarray}
\label{YCs_u_v-2NFs-final} \langle \, {\bf u}\, \alpha \,
|\psi_{1}\rangle &=& \frac{1}{{E-\frac{u_{1}^{2}}{m}
-\frac{3u_{2}^{2}}{4m}-\frac{2u_{3}^{2}}{3m}}}  \nonumber
\\ &\times&
 \, \int d^{3}u_{2}' \, \sum_{\gamma',\gamma'''} \,
g_{\alpha \gamma'''} \, \delta_{m'''_{s_{4}} m'_{s_{4}}}
\delta_{m'''_{s_{3}} m'_{s_{1}}} \, \delta_{m'''_{t_{4}} m'_{t_{4}}}
\delta_{m'''_{t_{3}} m'_{t_{1}}} \nonumber
\\  &\times& \,\, _{a}\langle{\bf u}_{1}\, m'''_{s_{1}}
m'''_{s_{2}} \, m'''_{t_{1}} m'''_{t_{2}} |t(\epsilon)
|\frac{-1}{2}{\bf u}_{2} -{\bf u}'_{2} \, m'_{s_{2}} m'_{s_{3}} \,
m'_{t_{2}} m'_{t_{3}} \rangle_{a} \, \nonumber
\\ &\times& \Biggl\{\,\, \sum_{\alpha''} \, g_{\gamma' \alpha''} \langle{\bf u}_{2}+\frac{1}{2} {\bf u}'_{2}
\,\, {\bf u}'_{2}\,\,{\bf u}_{3} \, \alpha''|\psi_{1}\rangle
\nonumber
\\ \quad && \hspace{2mm} - \sum_{\alpha''} \, g_{\gamma'_{1243} \alpha''} \, \langle{\bf u}_{2}+\frac{1}{2}
{\bf u}'_{2} \,\, \frac{1}{3}{\bf u}'_{2}+ \frac{8}{9}{\bf u}_{3}
\,\, {\bf u}'_{2}-\frac{1}{3}{\bf u}_{3} \, \alpha''
|\psi_{1}\rangle
 \nonumber \\  && \hspace{2mm} + \, \sum_{\beta'} \, g_{\gamma'
 \beta'} \,
 \langle{\bf u}_{2}+\frac{1}{2} {\bf u}'_{2}\,\, -{\bf u}'_{2}-\frac{2}{3}{\bf u}_{3} \,\,
\frac{1}{2}{\bf u}'_{2}-\frac{2}{3}{\bf u}_{3} \, \beta'
|\psi_{2}\rangle\,\, \Biggr\}, \nonumber \\
 \nonumber \\ \nonumber \\
\langle \, {\bf v}\, \beta \, |\psi_{2}\rangle &=&
\frac{\frac{1}{2}}{E-\frac{v_{1}^{2}}{m}
-\frac{v_{2}^{2}}{2m}-\frac{v_{3}^{2}}{m}} \nonumber \\ &\times&
 \int d^{3}v_{3}' \, \, \sum_{\gamma',\gamma'''} \,
g_{\beta \gamma'''} \, \delta_{m'''_{s_{3}} m'_{s_{1}}} \,
\delta_{m'''_{s_{4}} m'_{s_{2}}} \, \delta_{m'''_{t_{3}}
m'_{t_{1}}} \, \delta_{m'''_{t_{4}} m'_{t_{2}}} \nonumber \\
&\times& \,\, _{a}\langle {\bf v}_{1} \, m'''_{s_{1}} m'''_{s_{2}}
\, m'''_{t_{1}} m'''_{t_{2}} |t(\epsilon^{*})| {\bf v}'_{3}
 \, m'_{s_{3}} m'_{s_{4}} \, m'_{t_{3}} m'_{t_{4}} \rangle _{a}\, \nonumber
\\ &\times& \Biggl \{\,\, \sum_{\alpha'} \, g_{\gamma' \alpha'} \, \langle{\bf v}_{3}\,\,
 \frac{2}{3}{\bf v}_{2}+\frac{2}{3}{\bf v}'_{3} \,\, \frac{1}{2}{\bf v}_{2}-{\bf v}'_{3} \, \alpha' |\psi_{1}\rangle
  \nonumber \\ && \hspace{2mm} - \sum_{\alpha'} \, g_{\gamma'_{1243} \alpha'} \, \langle{\bf v}_{3}\,\,
 \frac{2}{3}{\bf v}_{2}-\frac{2}{3}{\bf v}'_{3} \,\, \frac{1}{2}{\bf v}_{2}+{\bf v}'_{3} \, \alpha' |\psi_{1}\rangle
\nonumber \\
  && \hspace{2mm}+ \sum_{\beta'} \, g_{\gamma' \beta'} \,
\langle{\bf v}_{3}\,\,-{\bf v}_{2}\,\, {\bf v}'_{3} \, \beta'
|\psi_{2}\rangle \,\, \Biggr\}.
 \end{eqnarray}

It is worth to be mentioned that if we ignore the spin-isospin
quantum numbers we can easily reach the bosonic type of
three dimensional Yakubovsky integral equations which are given in
Ref. \cite{Hadizadeh_FBS40}.

\section{Yakubovsky Equations in 3D Representation With 3NFs} \label{sec:YC_with_3NFs}
In this section we present the less cumbersome 3D representation of
the Yakubovsky components with the 3NFs in compare with PW
representation. The evaluation of the Yakubovsky components with the
inclusion of 3NFs will be exactly the same as Eq.
(\ref{YCs_u_v-2NFs-final}) except that an extra term with
$W_{123}^{(3)}$ occurs in the first component. This is
\begin{eqnarray}  \label{Eq.3NF-term}
 \langle \, {\bf u}\, \alpha \,| (1+G_{0}t) G_{0} V_{123}^{(3)} |\Psi\rangle
   &=& \frac{1}{{E-\frac{u_{1}^{2}}{m}
-\frac{3u_{2}^{2}}{4m}-\frac{2u_{3}^{2}}{3m}}} \nonumber \\
&\times&
 \Biggl\{\,  \langle \, {\bf u}\, \alpha \,|
V_{123}^{(3)} |\Psi\rangle + \langle \, {\bf u}\, \alpha \,| t G_{0}
V_{123}^{(3)} |\Psi\rangle \, \Biggr\}.
\end{eqnarray}

The matrix elements of the second term can be evaluated by inserting
the suitable completeness relations as:
\begin{eqnarray}
\label{Eq.3NF-completeness}  \langle \, {\bf u}\, \alpha \,| t G_{0}
V_{123}^{(3)} |\Psi\rangle &=&
 \sum_{\gamma'} \, \sum_{\gamma''} \, \intsum_{\alpha'''}^{\bf u'''}  \nonumber \\ &\times& \langle \,  \alpha \,| \, \gamma' \, \rangle
  \langle \, {\bf u} \, \gamma' \,|t G_{0} | \,
{\bf u}''' \, \gamma'' \, \rangle  \langle \,  \gamma'' \,| \,
\alpha''' \, \rangle  \langle {\bf u}''' \, \alpha'''\,
|V_{123}^{(3)} |\Psi\rangle \nonumber
\\ &\equiv&  \sum_{\gamma'} \, \sum_{\gamma''} \, \intsum_{\alpha'''}^{\bf u'''} \frac{g_{\gamma' \alpha}
\, g_{\gamma'' \alpha'''}}{{E-\frac{u_{1}'''^{2}}{m}
-\frac{3u_{2}'''^{2}}{4m}-\frac{2u_{3}'''^{2}}{3m}}}
 \nonumber \\ &\times& \langle \, {\bf u} \, \gamma' \,|t  | \,
{\bf u}''' \, \gamma'' \, \rangle    \langle {\bf u}''' \,
\alpha'''\, |V_{123}^{(3)} |\Psi\rangle,
\end{eqnarray}
after evaluating the matrix elements of the NN $t$-matrix, Eq.
(\ref{t_in_u}), and integrating over ${\bf u'''_{2}}$ and ${\bf
u'''_{3}}$ vectors, we obtain:
\begin{eqnarray}
\label{Eq.3NF-t-evaluated}
 \langle \,
{\bf u}\, \alpha \,| t G_{0} V_{123}^{(3)} |\Psi\rangle &=&
    \sum_{\gamma',\gamma'',\alpha'''} \,  \int d^{3} u'''_{1} \frac{g_{\gamma' \alpha}
\, g_{\gamma'' \alpha'''}}{{E-\frac{u_{1}'''^{2}}{m}
-\frac{3u_{2}^{2}}{4m}-\frac{2u_{3}^{2}}{3m}}} \nonumber \\
&\times& \delta_{m'_{s_{3}} m''_{s_{3}}} \delta_{m'_{s_{4}}
m''_{s_{4}}} \, \delta_{m'_{t_{3}} m''_{t_{3}}} \delta_{m'_{t_{4}}
m''_{t_{4}}} \nonumber
\\ &\times&
 \langle \, {\bf u}_{1} \, m'_{s_{1}}
m'_{s_{2}} m'_{t_{1}} m'_{t_{2}} |t(\epsilon) |{\bf u}'''_{1} \,
m''_{s_{1}} m''_{s_{2}} m''_{t_{1}} m''_{t_{2}} \rangle \nonumber
\\ &\times& \langle {\bf u}'''_{1} \, {\bf u}_{2} \, {\bf u}_{3} \,
\alpha'''\, |V_{123}^{(3)} |\Psi\rangle,
\end{eqnarray}

By using Eq. (\ref{t_in_u}), the symmetry property of the 3NF and
the anti-symmetry property of the total wave function under exchange
of nucleons 1 and 2, one can write:
\begin{eqnarray}
\label{Eq.3NF-t-under-p}  \langle \, {\bf u}\, \alpha \,| t G_{0}
V_{123}^{(3)} |\Psi\rangle &=&
    \sum_{\gamma',\gamma'',\alpha'''} \,  \int d^{3} u'''_{1} \frac{g_{\gamma' \alpha}
\, g_{\gamma'' \alpha'''}}{{E-\frac{u_{1}'''^{2}}{m}
-\frac{3u_{2}^{2}}{4m}-\frac{2u_{3}^{2}}{3m}}} \nonumber \\
&\times& \delta_{m'_{s_{3}} m''_{s_{3}}} \delta_{m'_{s_{4}}
m''_{s_{4}}} \, \delta_{m'_{t_{3}} m''_{t_{3}}} \delta_{m'_{t_{4}}
m''_{t_{4}}} \nonumber
\\ &\times&
 \langle \, {\bf u}_{1} \, m'_{s_{1}}
m'_{s_{2}} m'_{t_{1}} m'_{t_{2}} |t(\epsilon) P_{12} |-{\bf
u}'''_{1} \, m''_{s_{2}} m''_{s_{1}} m''_{t_{2}} m''_{t_{1}}
\rangle \nonumber \\ &\times& \left(- (-)^{s_{12}'''+t_{12}'''}
\right) \langle -{\bf u}'''_{1} \, {\bf u}_{2} \, {\bf u}_{3} \,
\alpha'''\, |V_{123}^{(3)} |\Psi\rangle,
\end{eqnarray}

under the exchange of the labels $m''_{s_{1}}, m''_{t_{1}}$ to
$m''_{s_{2}}, m''_{t_{2}}$, reverse of it and changing ${\bf
u}'''_{1}$ to $-{\bf u}'''_{1}$ we find:
\begin{eqnarray}
\label{Eq.3NF-exchange-lables}  \langle \, {\bf u}\, \alpha \,| t
G_{0} V_{123}^{(3)} |\Psi\rangle &=&
    \sum_{\gamma',\gamma'',\alpha'''} \,   \int d^{3} u'''_{1} \frac{g_{\gamma' \alpha}
\, g_{\gamma'' \alpha'''}}{{E-\frac{u_{1}'''^{2}}{m}
-\frac{3u_{2}^{2}}{4m}-\frac{2u_{3}^{2}}{3m}}} \nonumber \\
&\times& \delta_{m'_{s_{3}} m''_{s_{3}}} \delta_{m'_{s_{4}}
m''_{s_{4}}} \, \delta_{m'_{t_{3}} m''_{t_{3}}} \delta_{m'_{t_{4}}
m''_{t_{4}}} \nonumber
\\ &\times&
 \langle \, {\bf u}_{1} \, m'_{s_{1}}
m'_{s_{2}} m'_{t_{1}} m'_{t_{2}} |-t(\epsilon) P_{12} |{\bf
u}'''_{1} \, m''_{s_{1}} m''_{s_{2}} m''_{t_{1}} m''_{t_{2}}
\rangle \nonumber \\ &\times& \langle {\bf u}'''_{1} \, {\bf
u}_{2} \, {\bf u}_{3} \, \alpha'''\, |V_{123}^{(3)} |\Psi\rangle.
\end{eqnarray}

Now we consider Eqs. (\ref{Eq.3NF-t-evaluated}) and
 (\ref{Eq.3NF-exchange-lables}) together to achieve:
\begin{eqnarray}
\label{Eq.3NF-add} \langle \, {\bf u}\, \alpha \,| t G_{0}
V_{123}^{(3)} |\Psi\rangle &=& \frac{1}{2}
    \sum_{\gamma',\gamma'',\alpha'''} \,   \int d^{3} u'''_{1} \frac{g_{\gamma' \alpha}
\, g_{\gamma'' \alpha'''}}{{E-\frac{u_{1}'''^{2}}{m}
-\frac{3u_{2}^{2}}{4m}-\frac{2u_{3}^{2}}{3m}}} \nonumber \\
&\times& \delta_{m'_{s_{3}} m''_{s_{3}}} \delta_{m'_{s_{4}}
m''_{s_{4}}} \, \delta_{m'_{t_{3}} m''_{t_{3}}} \delta_{m'_{t_{4}}
m''_{t_{4}}} \nonumber
\\ &\times&  \langle \, {\bf u}_{1} \, m'_{s_{1}}
m'_{s_{2}} m'_{t_{1}} m'_{t_{2}} |t(\epsilon) (1-P_{12}) |{\bf
u}'''_{1} \, m''_{s_{1}} m''_{s_{2}} m''_{t_{1}} m''_{t_{2}}
\rangle \nonumber \\ &\times& \langle {\bf u}'''_{1} \, {\bf
u}_{2} \, {\bf u}_{3} \, \alpha'''\, |V_{123}^{(3)} |\Psi\rangle,
\end{eqnarray}
by applying the introduction of the anti-symmetrized NN $t$-matrix,
Eq. (\ref{eq.ta}), we can rewrite the final representation of the
three dimensional Yakubovsky integral equations, Eq.
(\ref{YCs_u_v-2NFs-final}), as:
\begin{eqnarray}
\label{YCs_u_v} \langle \, {\bf u}\, \alpha \, |\psi_{1}\rangle &=&
\frac{1}{{E-\frac{u_{1}^{2}}{m}
-\frac{3u_{2}^{2}}{4m}-\frac{2u_{3}^{2}}{3m}}}  \nonumber
\\ && \hspace{-20mm} \times \Biggl [\,
  \int d^{3}u_{2}' \, \sum_{\gamma',\gamma'''} \,
g_{\alpha \gamma'''} \, \delta_{m'''_{s_{4}} m'_{s_{4}}}
\delta_{m'''_{s_{3}} m'_{s_{1}}} \, \delta_{m'''_{t_{4}} m'_{t_{4}}}
\delta_{m'''_{t_{3}} m'_{t_{1}}} \nonumber
\\  &\times& \,\, _{a}\langle{\bf u}_{1}\, m'''_{s_{1}}
m'''_{s_{2}} \, m'''_{t_{1}} m'''_{t_{2}} |t(\epsilon)
|\frac{-1}{2}{\bf u}_{2} -{\bf u}'_{2} \, m'_{s_{2}} m'_{s_{3}} \,
m'_{t_{2}} m'_{t_{3}} \rangle_{a} \, \nonumber
\\ &\times& \Biggl\{\,\, \sum_{\alpha''} \, g_{\gamma' \alpha''} \langle{\bf u}_{2}+\frac{1}{2} {\bf u}'_{2}
\,\, {\bf u}'_{2}\,\,{\bf u}_{3} \, \alpha''|\psi_{1}\rangle
\nonumber
\\ \quad && \hspace{2mm} - \sum_{\alpha''} \, g_{\gamma'_{1243} \alpha''} \, \langle{\bf u}_{2}+\frac{1}{2}
{\bf u}'_{2} \,\, \frac{1}{3}{\bf u}'_{2}+ \frac{8}{9}{\bf u}_{3}
\,\, {\bf u}'_{2}-\frac{1}{3}{\bf u}_{3} \, \alpha''
|\psi_{1}\rangle
 \nonumber \\  && \hspace{2mm} + \, \sum_{\beta'} \, g_{\gamma'
 \beta'} \,
 \langle{\bf u}_{2}+\frac{1}{2} {\bf u}'_{2}\,\, -{\bf u}'_{2}-\frac{2}{3}{\bf u}_{3} \,\,
\frac{1}{2}{\bf u}'_{2}-\frac{2}{3}{\bf u}_{3} \, \beta'
|\psi_{2}\rangle\,\, \Biggr\} \nonumber \\  && \hspace{-17mm} +
 \Biggl\{\,\, \langle \, {\bf u}\, \alpha \, |V_{123}^{(3)}
|\Psi\rangle \nonumber \\  && \hspace{-10mm} + \frac{1}{2}
\sum_{\gamma',\gamma'',\alpha'''} \,  g_{\alpha \gamma'} \,
g_{\gamma'' \alpha'''}  \int d^{3}u_{1}'\,
\frac{\delta_{m'_{s_{3}} m''_{s_{3}}} \delta_{m'_{s_{4}}
m''_{s_{4}}} \delta_{m'_{t_{3}} m''_{t_{3}}} \delta_{m'_{t_{4}}
m''_{t_{4}}} }{E-\frac{u_{1}'^{2}}{m}
-\frac{3u_{2}^{2}}{4m}-\frac{2u_{3}^{2}}{3m}} \nonumber \\ &&
\hspace{-10mm} \quad \times \, _{a}\langle{\bf u}_{1}\, m'_{s_{1}}
m'_{s_{2}} \, m'_{t_{1}} m'_{t_{2}} |t(\epsilon) |{\bf u}'_{1} \,
m''_{s_{1}} m''_{s_{2}} \, m''_{t_{1}} m''_{t_{1}} \rangle_{a}
\langle{\bf u}'_{1}\,{\bf u}_{2}\,{\bf u}_{3} \,
\alpha'''\,|V_{123}^{(3)} |\Psi\rangle \Biggr\} \, \, \Biggr],
 \nonumber \\ \nonumber \\ \nonumber \\
\langle \, {\bf v}\, \beta \, |\psi_{2}\rangle &=&
\frac{1}{E-\frac{v_{1}^{2}}{m}
-\frac{v_{2}^{2}}{2m}-\frac{v_{3}^{2}}{m}} \nonumber \\ &\times&
\int d^{3}v_{3}' \, \, \sum_{\gamma',\gamma'''} \, g_{\beta
\gamma'''} \, \delta_{m'''_{s_{3}} m'_{s_{1}}} \,
\delta_{m'''_{s_{4}} m'_{s_{2}}} \, \delta_{m'''_{t_{3}}
m'_{t_{1}}} \, \delta_{m'''_{t_{4}} m'_{t_{2}}} \nonumber \\
&\times& \,\, _{a}\langle {\bf v}_{1} \, m'''_{s_{1}} m'''_{s_{2}}
\, m'''_{t_{1}} m'''_{t_{2}} |t(\epsilon^{*})| {\bf v}'_{3}
 \, m'_{s_{3}} m'_{s_{4}} \, m'_{t_{3}} m'_{t_{4}} \rangle _{a}\, \nonumber
\\ &\times& \Biggl \{\,\, \sum_{\alpha'} \, g_{\gamma' \alpha'} \, \langle{\bf v}_{3}\,\,
 \frac{2}{3}{\bf v}_{2}+\frac{2}{3}{\bf v}'_{3} \,\, \frac{1}{2}{\bf v}_{2}-{\bf v}'_{3} \, \alpha' |\psi_{1}\rangle
  \nonumber \\ && \hspace{2mm} - \sum_{\alpha'} \, g_{\gamma'_{1243} \alpha'} \, \langle{\bf v}_{3}\,\,
 \frac{2}{3}{\bf v}_{2}-\frac{2}{3}{\bf v}'_{3} \,\, \frac{1}{2}{\bf v}_{2}+{\bf v}'_{3} \, \alpha' |\psi_{1}\rangle
\nonumber \\
  && \hspace{2mm}+ \sum_{\beta'} \, g_{\gamma' \beta'} \,
\langle{\bf v}_{3}\,\,-{\bf v}_{2}\,\, {\bf v}'_{3} \, \beta'
|\psi_{2}\rangle \,\, \Biggr\}.
 \end{eqnarray}

To represent the generality of our 3D formalism we can simplify the
Eq. (\ref{YCs_u_v}) to the bosonic case by switching off the
spin-isospin quantum numbers, see Ref. \cite{Hadizadeh-EPJA36}. In
order to show the efficiency of our formalism we have also chosen a
realistic $2\pi$-exchange 3NF, i.e. TM, to evaluate the matrix
elements of $\langle \, {\bf u}\, \alpha \, |V_{123}^{(3)}
|\Psi\rangle$. As a matter of reference we present this formalism in
appendix (\ref{appendix:TM}) to indicate the simplicity of 3D
representation.

\section{The Number of Coupled Yakubovsky Equations in both the PW and the 3D Formalisms}\label{sec:number_of_YEs}

In this section we discuss about the number of coupled Yakubovsky
equations in both the 3D and the PW approaches. Already in the PW
representation of the 3N bound state equations the number of
channels must be high, e.g. of the order of 34, in order to achieve
reasonably well enough converged energy eigenvalues
\cite{Machleidt-ANP19}. In contrast to the 3N system the number of
channels for the 4N bound state $N=N_{\alpha} + N_{\beta}$, where
$N_{\alpha}$ and $N_{\beta}$ are the numbers of $\alpha$ and $\beta$
quantum number combinations respectively, is in principle unlimited
even if the 2N interaction is assumed to act only in a certain 2N
states. Nevertheless the approach to that unlimited number can be
classified for instance in the following possibly useful manner. For
fixed $l_{4}$ and fixed total quantum numbers $J$ and $T$ the
numbers of $\alpha$-states is strictly finite once the 2N
interaction is assumed to act only up to $j_{12}^{max}$.
Correspondingly, the number of $\beta$-states is strictly finite
once $l_{2}$ is fixed and again the 2N interaction is assumed to be
zero in states for $j_{12}, j_{34}<j_{ij}^{max}$. We display
examples for those maximum $N_{\alpha}$ and $N_{\beta}$ values in
Table \ref{Table_pw_channels}. Thus even if assuming only
$j_{ij}^{max}=1$, and restricting $l_{2}$ and $l_{4}$ to be at most
1, one leads to $N=60$ channels, while the total isospin is
restricted to be zero \cite{Kamada_NPA548}.

\begin{table}[] \label{Table_pw_channels}
\caption {The number of partial wave channels contributing to both
kind of the 4N Jacobi coordinates for $J^{\pi}=0^{+}$ dependent on
maximal values of intercluster orbital angular momenta $l_{i}$ and
total two-body angular momenta $j_{ij}$. For (a) and (b) parts the
total isospin is restricted to $T=0$ and for (c) part $T=0,1,2$. $N$
is the total number of channels, where $N_{\alpha}$ and $N_{\beta}$
are the number of channels corresponding to $3+1$ and $2+2$
partitions. The results of part (a) are given according to the
notation of Ref. \cite{Kamada_NPA548}, where $j_{12}, j_{34} \leq
j_{ij}^{max}$ and $l_{2}, l_{4} \leq l_{i}^{max}$, and the results
of parts (b) and (c) are given according to the notation of Ref.
\cite{Nogga_PRC65}, where $j_{12},j_{34}\leq j_{ij}^{max}$,
$l_{2},l_{3}, l_{4} \leq l_{i}^{max}$ and $l_{12}+l_{3}+l_{4},
l_{12}+l_{34}+l_{2} \leq l_{sum}^{max}$. It should be mentioned that
for (c) part, $N_{\alpha}=4200$ and $N_{\beta}=2000$.}
\begin{center}
\begin{tabular}{cccccccccccccccc} \hline
\hline
(a) Ref. \cite{Kamada_NPA548}  \\
& $j_{ij}^{max}$&  $N_{\alpha}$ & $N_{\beta}$ & $N$ \\
\hline
 $l_{i}^{max}=0$ && \\
& $1$ &  10 & 10 & 20  \\
& $2$ &  18 & 18 & 36  \\
& $3$ &  26 & 26 & 52 \\
\hline
 $l_{i}^{max}=1$ \\
& $1$ &  34 & 26 & 60\\
& $2$ &  66 & 58 & 124\\
& $3$ &  98 & 90 & 188\\
 $l_{i}^{max}=2$  \\
& $1$ &  62 & 34 & 96 \\
& $2$ &  130 & 98 & 228 \\
& $3$ &  202 & 170 & 372 \\
\hline
 $l_{i}^{max}=3$ \\
& $1$ &  90 & 34 & 124\\
& $2$ &  198 & 122 & 320\\
& $3$ &  322 & 242 & 564\\
\hline
 $l_{i}^{max}=4$  \\
& $1$ &  118 & 34 & 152  \\
& $2$ &  266 & 130 & 396   \\
& $3$ &  446 & 290 & 736  \\
  \\
  (b)  Ref. \cite{Kamada_PRC64}  \\
   $j_{ij}^{max}=6$ & $l_{i}^{max}=8$  & $l_{sum}^{max}=12$ & $T=0$ &  $N=1572$  \\
 \\
  (c)  Ref. \cite{Nogga_PRC65}  \\
   $j_{ij}^{max}=6$ & $l_{i}^{max}=8$  & $l_{sum}^{max}=14$ & $T=0,1,2$ & $N=6200$  \\
 \\
\hline \hline
\end{tabular}
\end{center}
\end{table}
\

\begin{table}[]
\caption {The number of the spin-isospin states for both kind of the
4N Jacobi coordinates in a realistic 3D formalism. $N_{\alpha_{S}} (
N_{\alpha_{T}})$ and $N_{\beta_{S}} (N_{\beta_{T}})$ are the number
of the spin (isospin) states for $3+1$ and $2+2$ partitions
correspondingly. }
\begin{center}
\begin{tabular}{cccccccccccccccccccc}
\hline \hline \\
 (a) $3+1$ partitions \\ \\
$((s_{12} \,\, \frac{1}{2})s_{123} \,\, \frac{1}{2}) S M_{S}$ &&
$S=0$ && $S=0,1$ && $S=0,1,2$  \\
\hline $((0 \,\, \frac{1}{2})\frac{1}{2} \,\, \frac{1}{2}) 0$ && 1
&& 1+0 && 1+0+0 \\
$((0 \,\, \frac{1}{2})\frac{1}{2} \,\, \frac{1}{2}) 1$ && 0 && 0+3
&& 0+3+0 \\
$((1 \,\, \frac{1}{2})\frac{1}{2} \,\, \frac{1}{2}) 0$ && 1 && 1+0
&& 1+0+0 \\
$((1 \,\, \frac{1}{2})\frac{1}{2} \,\, \frac{1}{2}) 1$ && 0 && 0+3
&& 0+3+0  \\
$((1 \,\, \frac{1}{2})\frac{3}{2} \,\, \frac{1}{2}) 1$ && 0 && 0+3
&& 0+3+0 \\
$((1 \,\, \frac{1}{2})\frac{3}{2} \,\, \frac{1}{2}) 2$ && 0 && 0+0
&& 0+0+5 \\ \hline
$N_{\alpha_{S}}$ && 2 && 11 && 16 \\ \\
$((t_{12} \,\, \frac{1}{2})t_{123} \,\, \frac{1}{2}) T 0$ && $T=0$
&& $T=0,1$ && $T=0,1,2$ \\ \hline $((0 \,\, \frac{1}{2})\frac{1}{2}
\,\, \frac{1}{2}) 0$ && 1 && 1+0 && 1+0+0 \\
$((0 \,\, \frac{1}{2})\frac{1}{2} \,\, \frac{1}{2}) 1$ && 0 &&
0+1 && 0+1+0 \\
$((1 \,\, \frac{1}{2})\frac{1}{2} \,\, \frac{1}{2}) 0$ && 1 &&
1+0 && 1+0+0 \\
$((1 \,\, \frac{1}{2})\frac{1}{2} \,\, \frac{1}{2}) 1$ && 0 && 0+1
&& 0+1+0 \\
$((1 \,\, \frac{1}{2})\frac{3}{2} \,\, \frac{1}{2}) 1$ && 0 && 0+1 && 0+1+0 \\
$((1 \,\, \frac{1}{2})\frac{3}{2} \,\, \frac{1}{2}) 2$ && 0 && 0+0
&& 0+0+1 \\ \hline
$N_{\alpha_{T}}$ && 2 && 5 && 6 \\
\\
(b) $2+2$ partitions \\ \\
$(s_{12}\,\, s_{34}) S M_{S}$ && $S=0$ && $S=0,1$ && $S=0,1,2$  \\

\hline $(0 \,\, 0) 0 $ && 1 && 1+0 && 1+0+0  \\
$(0 \,\, 1) 1 $ && 0 && 0+3 && 0+3+0  \\
$(1 \,\, 0) 1 $ && 0 && 0+3 && 0+3+0  \\
$(1 \,\, 1) 0 $ && 1 && 1+0 && 1+0+0  \\
$(1 \,\, 1) 2 $ && 0 && 0+0 && 0+0+5 \\ \hline
$N_{\beta_{S}}$ && 2 && 8 && 13  \\ \\
$(t_{12}\,\, t_{34}) T 0$ && $T=0$ && $T=0,1$ && $T=0,1,2$ \\ \hline
$(0 \,\, 0) 0 $ && 1 && 1+0 && 1+0+0 \\
$(0 \,\, 1) 1 $ && 0 && 0+1 && 0+1+0 \\
$(1 \,\, 0) 1 $ && 0 && 0+1 && 0+1+0 \\
$(1 \,\, 1) 0 $ && 1 && 1+0 && 1+0+0 \\
$(1 \,\, 1) 2 $ && 0 && 0+0 && 0+0+1 \\ \hline
$N_{\beta_{T}}$ && 2 && 4 && 5 \\
\\
 \hline\hline
\end{tabular}
\end{center}
\label{Table_3D_states}
\end{table}

\begin{table}[]
\caption {The number of coupled Yakubovsky equations for $\alpha$
particle in a realistic 3D formalism according to the spin-isospin
states $(S-T)$ that we have taken into account. $N=N_{\alpha} +
N_{\beta}$ is the total number of coupled Yakubovsky equations,
where $N_{\alpha}=N_{\alpha_{S}}\times N_{\alpha_{T}}$ and
$N_{\beta}=N_{\beta_{S}}\times N_{\beta_{T}}$ are the number of
$3+1$ and $2+2$ states correspondingly. The star superscript
indicates all spin or isospin states that we have taken into account
up to a specific value. }
\begin{center}
\begin{tabular}{cccccccccccccccccccc}
\hline \hline $(S-T)$ && $N_{\alpha}$ && $N_{\beta}$ && $N$ \\\hline
$(0-0)$                && 4   && 4   && 8 \\
$(1^{\ast}-0)$         && 22  && 16  && 38  \\
$(2^{\ast}-0)$         && 32  && 26  && 58 \\
$(0-1^{\ast})$         && 10  && 8   &&  18\\
$(1^{\ast}-1^{\ast})$  && 55  && 32  && 87 \\
$(2^{\ast}-1^{\ast})$  && 80  && 52  && 132 \\
$(0-2^{\ast})$         && 12  && 10  && 22\\
$(1^{\ast}-2^{\ast})$  && 66  && 40  && 106 \\
$(2^{\ast}-2^{\ast})$  && 96  && 65  && 161 \\
\hline\hline
\end{tabular}
\end{center}
\label{Table_3D_Ystates}
\end{table}

In Tables \ref{Table_3D_states} and \ref{Table_3D_Ystates} we
present the number of the spin-isospin states for both kind of the
Jacobi coordinates, $\alpha$ and $\beta$, as well as the number of
coupled Yakubovsky equations in our realistic 3D formalism. Since
the angular momentum quantum numbers, i.e. $l_{12},l_{3},l_{4}$ and
$l_{12},l_{2},l_{34}$, do not appear explicitly in our formalism,
therefore the number of coupled equations which are fixed according
to the spin-isospin states are strongly reduced. This is an
indication that the present formalism automatically consider all
partial waves without any truncation on the space part. Considering
the spin-isospin degrees of freedom for $^{4}He$ one leads to 8, 38,
58, 18, 87, 132, 22, 106 and 161 coupled equations for different
combinations of total spin-isospin states $(S-T)$: $(0-0)$,
$(1^{\ast}-0)$, $(2^{\ast}-0)$, $(0-1^{\ast})$,
$(1^{\ast}-1^{\ast})$, $(2^{\ast}-1^{\ast})$, $(0-2^{\ast})$,
$(1^{\ast}-2^{\ast})$ and $(2^{\ast}-2^{\ast})$ respectively. The
star superscript indicates all spin or isospin states that we have
taken into account up to a specific value. It is clear that in the
3D formalism for a fully charge dependent calculation there is only
161 coupled equations, whereas in PW approach after truncation of
the Hilbert space to $T=0$ there is 1572 coupled equation, see part
(b) of Table \ref{Table_pw_channels}, and 6200 coupled equations for
a fully charge dependent, see part (c) of Table
\ref{Table_pw_channels}. So our 3D formalism leads to a very small
number of coupled equations in comparison with the very large number
of coupled equations in the truncated PW formalism.

However, it should be mentioned that our formulation leads to
coupled integral equations which depend on three vector variables
for the amplitudes, whereas the PW formulation after truncation
leads to a finite number of coupled equations in three variables
for the amplitudes.

The coupled Yakubovsky equations, Eq. (\ref{YCs_u_v}), represent a
set of three dimensional homogenous integral equations, which
after discretization turns into a huge matrix eigenvalue
equation. In order to solve these coupled integral equations
directly without employing the PW projections, one has to define a
coordinate system. According to experience of $^{3}$H binding
energy calculations, Ref. \cite{Bayegan-PRC77}, it is convenient
to choose the spin polarization direction parallel to the $z$-axis
and express the momentum vectors in this coordinate system. As
indicated in Ref. \cite{Hadizadeh-EPJA36} generally one needs six independent
variables, three magnitudes of the Jacobi momentum vectors and
three angle variables, to uniquely specify the geometry of the
three vectors. But in contrast to the bosonic case, Ref.
\cite{Hadizadeh_FBS40}, one does not have this freedom to choose
the third vectors, i.e. ${\bf u}_{3}$ and ${\bf v}_{3}$, parallel
to the $z-$ axis and second vectors, i.e. ${\bf u}_{2}$ and ${\bf
v}_{2}$, in the $x-z$ plane. So the angular dependencies of
spin-isospin-dependent Yakubovsky components will be more
complicated rather than bosonic ones. The dependence on the
continuous momentum and angle variables should be replaced in the
numerical treatment by a dependence on certain discrete values.
Let the number of these discrete points be denoted by $N_{jacobi}$
and $N_{angle}$ corresponding to momentum and angle variables, the
dimension of the eigenvalue problem in both PW and 3D approaches
is:
\begin{eqnarray}
N^{PW} &=& N_{jacobi}^{3}\times (N_{\alpha}^{PW}
+ N_{\beta}^{PW}) \nonumber \\
N^{3D} &=& N_{jacobi}^{3}\times N_{angle}^{3}\times
(N_{\alpha}^{3D} + N_{\beta}^{3D})
 \label{eq.dimension}
\end{eqnarray}

It should be indicated that in the formulation of the 4N bound state
we can consider the parity and the time reversal invariance to cut
down the dimension of the Yakubovsky components, see appendix
(\ref{appendix:Parity_Time-reversal}). So according to our
experience in four-bosonic calculations and also by considering
the parity and the time reversal invariance one needs in 4N
realistic calculations thirty grid points for Jacobi momentum
variables and ten grid points for angle variables. By these number
of mesh grids the dimension of the eigenvalue problem for a fully
charge dependent calculation, i.e. for total spin-isospin states
$(S^{\ast}-T^{\ast})=(2^{\ast}-2^{\ast})$, will be:
\begin{eqnarray}
N^{PW} &=& 30^{3}\times (4200 + 2000) \cong 1.7 \times 10^8 \nonumber \\
N^{3D} &=& 30^{3}\times 10^{3}\times (96 + 65) \cong 4.4 \times
10^9
 \label{eq.dimension-order}
\end{eqnarray}
Thus, the price for the smaller number of equations in 3D
representation is the higher dimensionality of the integral
equations. In other words, algebraic simplification is achieved by
a more involved numerical scheme. As indicated in ref
\cite{Hadizadeh_FBS40} for bosonic case these dimensions reduce to
the  values $N^{PW}\cong 2.7 \times 10^6$ and $N^{3D}\cong 2.7
\times 10^7$. As it is clear from the close similarity of
dimension values and also based on the experience of simpler
bosonic calculations, we expect that for more rigorous nucleonic
case the numerical calculation could be achievable.

\section{Summary and Outlook}\label{sec:summary}

We propose a new representation of three dimensional Yakubovsky
equations for the four-nucleon bound state including the spin and
isospin degrees of freedom in the momentum space. This formalism
stays closely to the bosonic structure where the spin and isospin
degrees of freedom are ignored. This is an important step forward
since the formulation based on partial wave decomposition, which
includes the spin and isospin degrees of freedom, after truncation
leads to two coupled sets of a finite number of coupled equations in
three variables for the amplitudes. In contrast our 3D formulation
leads to two coupled sets of a strictly finite number of equations
in three vector variables for the amplitudes. The comparison of 3D
and PW formalisms shows that our 3D formalism avoids the very
involved angular momentum algebra occurring for the permutations and
transformations and it is more efficient and less cumbersome for
considering the three-nucleon forces. This formalism enables us to
handle realistic 2N and 3N potentials with all their complexity in
four-nucleon bound state calculations. This work provides the
necessary formalism for the calculation of four-nucleon binding
energy which is under preparation.

\section*{Acknowledgments}
This work was supported by the research council of the University of
Tehran.

\appendix

\section{The Evaluation of $\langle \, {\bf u}\, \alpha \, |V_{123}^{(3)} |\Psi\rangle$ for the TM $2\pi$-exchange 3NF }
\label{appendix:TM}

\subsection{Preparation of the TM 3NF for the 3D Representation} \label{subsec:TM_NN-like}

For the evaluation of the coupled Yakubovsky equations, Eq.
(\ref{YCs_u_v}), matrix elements of the form $\langle \, {\bf u}\,
\alpha \, |V_{123}^{(3)} |\Psi\rangle$ need to be calculated. In
this section we evaluate these matrix elements in our 3D approach
for the TM $2\pi$-exchange 3NF. In the notation from
Fig.~\ref{fig_3NF}, this force is given by \cite{Coon-PRC23}
\begin{eqnarray}
\label{TM}
 \nonumber V_{123}^{(3)}&=& V_{0}
\, \sigma_{2}\cdot {\bf Q}'\ \sigma_{1}\cdot {\bf Q}\
{F({Q}^{2})\over  {Q}^{2}+m_{\pi}^{2}}\
{F({Q}'^{2})\over {Q}'^{2}+m_{\pi}^{2}}\\
&\times& \left[ \tau_{1}\cdot \tau_{2}\ \left( A+B\ {\bf Q} \cdot
{\bf Q}'+C\ ({Q}^{2}+{Q}'^{2})\right) +D\ \tau_{3}\cdot
\tau_{1}\times\tau_{2}\ \sigma_{3}\cdot {\bf Q}\times {\bf Q}'
\right], \nonumber \\
\end{eqnarray}
with the momentum transfers ${\bf Q}$ and ${\bf Q}'$. The
$\sigma_i$'s are Pauli spin matrices and the form factors are chosen
to be $ F(Q^{2})=({\Lambda^{2}-m_{\pi}^{2}\over
\Lambda^{2}+Q^{2}})^{2} $ with the cut-off parameter $\Lambda$. One
distinguishes four terms in the TM force, the so called $A$-, $B$-,
$C$- and $D$-term. The constants $A$, $B$, $C$ and $D$ are
determined by the low energy theorems \cite{Coon-NPA317}. The
momentum transfers are given by
\begin{eqnarray}
\label{momentum_transfers} {\bf Q}&=&{\bf k}_{1}-{\bf k}'_{1} \equiv
\left \{(+{\bf u}_{1}-\frac{1}{2}{\bf u}_{2})-(+{\bf
u}'_{1}-\frac{1}{2}{\bf u}'_{2})\right\}_{(123,4)}
\equiv \{{\bf u}_{2}-{\bf u}'_{2}\}_{(231,4)}, \nonumber \\
{\bf Q}'&=&{\bf k}_{2}'-{\bf k}_{2}  \equiv \left\{ (-{\bf
u}'_{1}-\frac{1}{2}{\bf u}'_{2})-(-{\bf u}_{1}-\frac{1}{2}{\bf
u}_{2})\right\}_{(123,4)} \equiv \{{\bf u}'_{2}-{\bf
u}\,_{2}\}_{(312,4)},
\end{eqnarray}
as indicated in Fig. \ref{fig_3NF}. In order to be able to evaluate
the TM 3NF matrix elements we should evaluate the scalar product of
the spin and momentum transfer vectors, $\sigma_{1}\cdot {\bf Q}$
and $\sigma_{2}\cdot {\bf Q}'$, the scalar product of both the
momentum transfer vectors, ${\bf Q} \cdot {\bf Q}'$, the scalar
triple product of the spin and momentum transfer vectors,
$\sigma_{3}\cdot {\bf Q}\times {\bf Q}'$, as well as the isospin
operators. Toward this aim we can rewrite the spin-space operators
in the form that can be evaluated easily in our 3D representation
as:
\begin{eqnarray}
\label{spin_momentum_operators} \sigma_{1}\cdot {\bf Q}&=& Q \
\sigma_{1}\cdot  \widehat{{\bf Q}},
\nonumber \\
\sigma_{2}\cdot {\bf Q}'&=& Q' \ \sigma_{2}\cdot \widehat{{\bf Q}}',
\nonumber \\
 {\bf Q} \cdot {\bf Q}' &=& Q Q' \gamma,
\nonumber \\
 \sigma_{3}\cdot {\bf Q}\times {\bf Q}' &=& |{\bf Q}\times {\bf Q}'| \  \sigma_{3}\cdot \widehat{{\bf Q}\times {\bf
 Q}'} \nonumber \\
 &=& Q Q' a \ \Biggl \{ \frac{3ia}{4\gamma} -\frac{i}{2 \gamma a} \ (\sigma_{3}\cdot \widehat{{\bf
 Q}}')^{2}+\frac{i}{2 a} \ \sigma_{3}\cdot \widehat{{\bf
 Q}}' \ \sigma_{3}\cdot \widehat{{\bf Q}} \nonumber \\ && \quad\quad\quad +\frac{i}{8 \gamma a} \ (\sigma_{3}\cdot \widehat{{\bf
 Q}}')^{2} \ (\sigma_{3}\cdot \widehat{{\bf Q}})^{2} - \frac{i}{2 \gamma a} \ (\sigma_{3}\cdot \widehat{{\bf
 Q}})^{2}  \Biggr \},
\end{eqnarray}
where
\begin{eqnarray}
  \label{gamma_a} \gamma= \widehat{{\bf Q}}.\widehat{{\bf Q}}' \ \ ,
  a=\sqrt{1-\gamma^{2}}.
\end{eqnarray}

The scalar product of the spin-momentum vectors can be evaluated as
\begin{eqnarray}
\label{s_dot_Q} \langle \, \hat{z} \, m'_{s} \,| \sigma \cdot
\widehat{{\bf Q}} | \, \hat{z}\, m''_{s} \, \rangle \ = \sum_{m_{s}}
m_{s} \ D_{m'_{s} m_{s}}^{\frac{1}{2}} (\widehat{{\bf Q}}) \ \
D_{m''_{s} m_{s}}^{\frac{1}{2}\,*} (\widehat{{\bf Q}}) = {\O}
_{m'_{s} m''_{s}} ^{ \widehat{{\bf Q}}},
\end{eqnarray}
where $D_{m'_{s} m_{s}}^{\frac{1}{2}}$ is Wigner D-function which is
defined generally as $D_{m'_{s} m_{s}}^{s} (\hat{q})= \langle
\hat{z} s m'_{s} | \hat{q} s m_{s} \rangle$ \cite{rose}. The application of the
TM 3NF to the total wave function $|\Psi\rangle$ can be considered
as sum of the the four independent terms:
\begin{eqnarray}
  \label{psi}
 | \psi \rangle= V_{123}^{(3)} |\Psi\rangle= \sum_{i=A}^{D} V_{31}^{(i)} \ I^{(i)} \ V_{23}^{(i)} | \Psi \rangle
 = \sum_{i=A}^{D}  | \psi^{i} \rangle,
 \end{eqnarray}
 where we have abbreviated the isospin
operators by $I^{(i)}$:
\begin{eqnarray}
\label{37}
 I^{(i)}=
  \left\{%
    \begin{array}{ll}
    \tau_{1}\cdot \tau_{2}\, & \hbox{i=A, B, C} \\
    \tau_{3}\cdot \tau_{1}\times\tau_{2}\, & \hbox{i=D}. \\
    \end{array}%
\right.
\end{eqnarray}

\subsection{Explicit 3D Evaluation of the A-, B-, C- and D-Terms} \label{sec:TM_abcd}

In the following we evaluate the matrix elements of $\langle \, {\bf
u}\, \alpha \, |\psi^{i}\rangle$ terms. From Fig.~\ref{fig_3NF} we
see that $V_{123}^{(3)}$ can be splitted into two parts and each
part contains a meson exchange. The first meson is exchanged in the
subsystem (31), where it is called for convenience subsystem 2, and
the second is exchanged in the subsystem (23), is called 1. Since
the structure of the 3BF is specified by two momentum transfers of
consecutive meson exchanges, it is convenient to insert a complete
set of states of the type $2$ between $V_{123}^{(3)}$ and
$|\Psi\rangle$ and another complete set of states of type $1$
between the two meson exchanges. Then the matrix element of
$V_{123}^{(3)}$ is rewritten as
\begin{eqnarray}
\label{psi_u} _{3}\langle \, {\bf u}\, \alpha \, |\psi^{i} \rangle
&=& \intsum_{\alpha'}^{\bf u'} \,\, _{3}\langle \, {\bf u}\,
\alpha \,|  \, {\bf u}'\, \alpha' \, \rangle_{1} \nonumber \\
&\times& \intsum_{\alpha''}^{\bf u''} \,\, _{1}\langle \, {\bf u}'\,
\alpha' \,|V_{31}^{(i)}| \, {\bf u}''\, \alpha'' \, \rangle_{1}
\nonumber
\\ &\times&
 \intsum_{\alpha'''}^{\bf u'''} \,\, _{1}\langle \, {\bf u}''\, \alpha''
 \,|I^{(i)}|
 \, {\bf u}'''\, \alpha''' \, \rangle_{2}
 \nonumber \\ &\times& \intsum_{\alpha''''}^{\bf u''''} \,\,
_{2}\langle \, {\bf u}'''\, \alpha''' \,
|V_{23}^{(i)}| \, {\bf u}''''\, \alpha'''' \,\rangle_{2}  \nonumber \\
&\times& \,_{2} \langle \, {\bf u}''''\, \alpha'''' \,|\Psi\rangle,
\end{eqnarray}
Here the subscripts $1, 2, 3$ of the bra and ket vectors stand for
the different types of three-nucleon coordinate systems of
$(3+1)$-type fragmentation $(ijk,4)$. The coordinate transformation
from the system of type 1 to one of type 3 can be evaluated
explicitly as:
\begin{eqnarray}
\label{u_1_to_3}
 _{3}\langle \, {\bf u}\, \alpha
\,|  \, {\bf u}'\, \alpha' \, \rangle_{1} =g_{\alpha_{3}
\alpha'_{1}} \, \delta^{3}({\bf u}'_{1}+\frac{1}{2}{\bf
u}_{1}+\frac{3}{4}{\bf u}_{2}) \delta^{3}({\bf u}'_{2}-{\bf
u}_{1}+\frac{1}{2}{\bf u}_{2}) \delta^{3}({\bf u}'_{3}-{\bf u}_{3}).
\end{eqnarray}

The matrix elements of the isospin coordinate transformations can be
rewritten as:
\begin{eqnarray}
\label{I_u}   _{1}\langle \, {\bf u}''\, \alpha''
 \,|I^{(i)}|
 \, {\bf u}'''\, \alpha''' \, \rangle_{2} = \ _{1}\langle \, {\bf u}''\, \alpha''_{S}
 \,| \, {\bf u}'''\, \alpha'''_{S} \, \rangle_{2} \ \  _{1}\langle \, \alpha''_{T}
 \,|I^{(i)}|
 \, \alpha'''_{T} \, \rangle_{2},
\end{eqnarray}
where
\begin{eqnarray}
\label{u_2_to_1}
  _{1}\langle \, {\bf u}''\, \alpha''_{S}
 \,| \, {\bf u}'''\, \alpha'''_{S} \, \rangle_{2} \ = g_{\alpha''_{1}
 \alpha'''_{2}}^{S} \,
 \delta^{3}({\bf u}'''_{1}+\frac{1}{2}{\bf
u}''_{1}+\frac{3}{4}{\bf u}''_{2} ) \delta^{3}({\bf u}'''_{2} -{\bf
u}''_{1}+\frac{1}{2}{\bf u}''_{2}) \delta^{3}({\bf u}'''_{3}-{\bf
u}''_{3}). \nonumber \\
\end{eqnarray}

The matrix elements of the coordinate transformations
 $_{1}\langle \, \alpha''_{T} \,|I^{(i)}|  \, \alpha'''_{T} \, \rangle_{2}$
 are given in Ref. \cite{Nogga_PhD}. The isospin matrix elements have been derived in
Ref. \cite{Huber-FBS22} for the 3N system and are given below for
the 4N system for completeness:
\begin{eqnarray}
\label{I^i}
  _{1}\langle \, \alpha''_{T}
 \,|I^{(i)}|
 \, \alpha'''_{T} \, \rangle_{2} =
  \left\{%
    \begin{array}{ll}
   \delta_{T''T'''} \delta_{M_T'' M_T'''} \delta_{t''_{231} t'''_{312}}
    (-6) \ (-)^{t''_{23}} \sqrt{ \hat t''_{23} \hat t'''_{31}}
       \\ \times \left\{
            \begin{array}{ccc}
              {1 \over 2 } & {1 \over 2 } &   t'''_{31}      \cr
              {1 \over 2 } &      1       &   {1 \over 2 } \cr
                t''_{23}     & {1 \over 2 } &   t''_{231}      \cr
            \end{array}
          \right\}  \, & \hbox{i=A, B, C} \\\\\\
     \delta_{T''T'''} \delta_{M_T'' M_T'''}
\delta_{t''_{231} t'''_{312} }
    24i  \ (-)^{2 t''_{231} } \sqrt{ \hat t''_{23} \hat t'''_{31}} \cr
       \times \sum_{\lambda=\frac{1}{2}}^{t''_{23}+ \frac{1}{2}} (-)^{3
\lambda + { 1 \over 2 }} \
       \ \left\{
            \begin{array}{ccc}
              \lambda      & {1 \over 2 } &   1            \cr
              {1 \over 2 } & {1 \over 2 } &  t''_{23}        \cr
            \end{array}
          \right\}
       \\ \times \left\{
            \begin{array}{ccc}
              t''_{231}      & {1 \over 2 } &   t''_{23}      \cr
              {1 \over 2 } &      1       & \lambda       \cr
                t'''_{31}    & {1 \over 2 } &  {1 \over 2 } \cr
            \end{array}
          \right\}  \, & \hbox{i=D}. \\
    \end{array}%
\right.
\end{eqnarray}

In the following steps the matrix elements for the different $V$'s
are evaluated. In the first step we evaluate the matrix elements of
$V_{31}^{i}$. Since both pion-exchange propagators in the 3NF term
only depend on the momentum transfer in a two-body subsystem, as
indicated in Eq.~(\ref{momentum_transfers}), and also because of the
separation of the isospin parts of 3NF, the matrix elements of
$V_{31}^{i}$ can be written as:
\begin{eqnarray}
\label{W_31^i} _{1}\langle \, {\bf u}'\, \alpha' \,|V_{31}^{(i)}| \,
{\bf u}''\, \alpha'' \, \rangle_{1} =   \delta^{3}({\bf u}'_{1}-{\bf
u}''_{1}) \ \delta^{3}({\bf u}'_{3}-{\bf u}''_{3}) \
\delta_{\alpha'_{T} \alpha''_{T}} \ _{1}\langle \, {\bf u}'_{2}\,
\alpha'_{S} \,|V_{31}^{(i)}| \, {\bf u}''_{2}\, \alpha''_{S} \,
\rangle_{1}, \nonumber \\
\end{eqnarray}
and the spin-space parts can be more simplified for the
$A$-,$B$-,$C$- and $D$-terms separately as:
\begin{eqnarray}
\label{W_31^ABC}
 _{1}\langle \, {\bf u}'_{2}\, \alpha'_{S}
\,|V_{31}^{(A, B, C)}| \, {\bf u}''_{2}\, \alpha''_{S} \,
\rangle_{1} &=& \sum_{\gamma'_{S}} \sum_{\gamma''_{S}}
g_{\alpha'_{S} \gamma'_{S}}^{1} \ g_{\alpha''_{S} \gamma''_{S}}^{1}
\ \delta_{m'_{s_{2}} m''_{s_{2}}} \delta_{m'_{s_{3}} m''_{s_{3}}}
\delta_{m'_{s_{4}} m''_{s_{4}}} \nonumber \\ &\times& \ _{1}\langle
\, {\bf u}'_{2}\, m'_{s_{1}} \,|V_{31}^{(A, B, C)}| \, {\bf
u}''_{2}\, m''_{s_{1}} \, \rangle_{1},
\end{eqnarray}
\begin{eqnarray}
\label{W_31^D}
 _{1}\langle \, {\bf u}'_{2}\, \alpha'_{S}
\,|V_{31}^{(D)}| \, {\bf u}''_{2}\, \alpha''_{S} \, \rangle_{1} &=&
\sum_{\gamma'_{S}} \sum_{\gamma''_{S}} g_{\alpha'_{S}
\gamma'_{S}}^{1} \ g_{\alpha''_{S} \gamma''_{S}}^{1} \
\delta_{m'_{s_{2}} m''_{s_{2}}} \delta_{m'_{s_{4}} m''_{s_{4}}}
\nonumber \\ &\times& \ _{1}\langle \, {\bf u}'_{2}\, m'_{s_{1}} \,
m'_{s_{3}} \,|V_{31}^{(D)}| \, {\bf u}''_{2}\, m''_{s_{1}} \,
m''_{s_{3}} \, \rangle_{1},
\end{eqnarray}
where the $A$-,$B$- and $C$-terms can be evaluated as:
\begin{eqnarray}
\label{W_31^A}
 _{1}\langle \, {\bf u}'_{2}\, m'_{s_{1}} \,|V_{31}^{(A)} | \, {\bf
u}''_{2}\, m''_{s_{1}} \, \rangle_{1} &=& \ _{1}\langle \, {\bf
u}'_{2}\, m'_{s_{1}} \,|F(Q^{2})\ {\sigma_{1}\cdot \widehat{{\bf
Q}}\over Q^{2}+m_{\pi}^{2}} \  Q| \, {\bf u}''_{2}\, m''_{s_{1}} \,
\rangle_{1} \nonumber \\ &=& \frac{F(({\bf u}'_{2}-{\bf
u}''_{2})^{2})}{({\bf u}'_{2}-{\bf u}''_{2})^{2}+m_{\pi}^{2}} \
|{\bf u}'_{2}-{\bf u}''_{2}| \ {\O} _{m'_{s_{1}} m''_{s_{1}}} ^{
\widehat{({\bf u}'_{2}-{\bf u}''_{2})} },
\end{eqnarray}
\begin{eqnarray}
\label{W_31^B}
 _{1}\langle \, {\bf u}'_{2}\, m'_{s_{1}} \,|V_{31}^{(B)} | \, {\bf
u}''_{2}\, m''_{s_{1}} \, \rangle_{1} &=& \ _{1}\langle \, {\bf
u}'_{2}\, m'_{s_{1}} \,|F(Q^{2})\ {\sigma_{1}\cdot \widehat{{\bf
Q}}\over Q^{2}+m_{\pi}^{2}} \  Q^{2} | \, {\bf u}''_{2}\,
m''_{s_{1}} \, \rangle_{1} \nonumber \\ &=& \frac{F(({\bf
u}'_{2}-{\bf u}''_{2})^{2})}{({\bf u}'_{2}-{\bf
u}''_{2})^{2}+m_{\pi}^{2}} \ |{\bf u}'_{2}-{\bf u}''_{2}|^{2} \ {\O}
_{m'_{s_{1}} m''_{s_{1}}} ^{ \widehat{({\bf u}'_{2}-{\bf
u}''_{2})}},
\end{eqnarray}
\begin{eqnarray}
\label{W_31^C}
 _{1}\langle \, {\bf u}'_{2}\, m'_{s_{1}} \,|V_{31}^{(C)} | \, {\bf
u}''_{2}\, m''_{s_{1}} \, \rangle_{1}&=& \ _{1}\langle \, {\bf
u}'_{2}\, m'_{s_{1}} \,| F(Q^{2})\ {\sigma_{1}\cdot \widehat{{\bf
Q}}\over Q^{2}+m_{\pi}^{2}} \ Q^{3} | \, {\bf u}''_{2}\, m''_{s_{1}}
\, \rangle_{1} \nonumber \\ &=& \frac{F(({\bf u}'_{2}-{\bf
u}''_{2})^{2})}{({\bf u}'_{2}-{\bf u}''_{2})^{2}+m_{\pi}^{2}} \
|{\bf u}'_{2}-{\bf u}''_{2}|^{3} \ {\O} _{m'_{s_{1}} m''_{s_{1}}} ^{
\widehat{({\bf u}'_{2}-{\bf u}''_{2})} },
\end{eqnarray}
and for the different parts of the $D$-term, see Eq.
(\ref{spin_momentum_operators}), we obtain:
\begin{eqnarray}
\label{W_31^D_1_2}
 _{1}\langle \, {\bf u}'_{2}\, m'_{s_{1}}
\, m'_{s_{3}} \,|V_{31}^{(D)} \ ^{1^{th}}| \, {\bf u}''_{2}\,
m''_{s_{1}} \, m''_{s_{3}} \, \rangle_{1} &=& \  _{1}\langle \, {\bf
u}'_{2}\, m'_{s_{1}} \, m'_{s_{3}} \,|V_{31}^{(D)} \ ^{2^{th}}| \,
{\bf u}''_{2}\, m''_{s_{1}} \, m''_{s_{3}} \, \rangle_{1} \nonumber
\\ &=& \delta _{m'_{s_{3}} m''_{s_{3}}} \ _{1}\langle \, {\bf
u}'_{2}\, m'_{s_{1}} \,|V_{31}^{(B)} | \, {\bf u}''_{2}\,
m''_{s_{1}} \, \rangle_{1},
\end{eqnarray}
\begin{eqnarray}
\label{W_31^D_3}
 && _{1}\langle \, {\bf u}'_{2}\, m'_{s_{1}}
\, m'_{s_{3}} \,|V_{31}^{(D)} \ ^{3^{th}}| \, {\bf u}''_{2}\,
m''_{s_{1}} \, m''_{s_{3}} \, \rangle_{1} \nonumber \\ && = \,
_{1}\langle \, {\bf u}'_{2}\, m'_{s_{1}} \, m'_{s_{3}} \,| F(Q^{2})\
{\sigma_{1}\cdot \widehat{{\bf Q}}\over Q^{2}+m_{\pi}^{2}} \ Q^{2} \
\sigma_{3}\cdot \widehat{{\bf Q}} | \, {\bf u}''_{2}\, m''_{s_{1}}
\, m''_{s_{3}} \, \rangle_{1} \nonumber \\ && = \, _{1}\langle \,
{\bf u}'_{2}\, m'_{s_{1}} \,|V_{31}^{(B)} | \, {\bf u}''_{2}\,
m''_{s_{1}} \, \rangle_{1} \ {\O} _{m'_{s_{3}} m''_{s_{3}}} ^{
\widehat{({\bf u}'_{2}-{\bf u}''_{2})} },
\end{eqnarray}
\begin{eqnarray}
\label{W_31^D_4_5} && _{1}\langle \, {\bf u}'_{2}\, m'_{s_{1}} \,
m'_{s_{3}} \,|V_{31}^{(D)} \ ^{4^{th}}| \, {\bf u}''_{2}\,
m''_{s_{1}} \, m''_{s_{3}} \, \rangle_{1} \nonumber \\ && = \,
_{1}\langle \, {\bf u}'_{2}\, m'_{s_{1}} \, m'_{s_{3}}
\,|V_{31}^{(D)} \ ^{5^{th}}| \, {\bf u}''_{2}\, m''_{s_{1}} \,
m''_{s_{3}} \, \rangle_{1} \nonumber \\ && = \, _{1}\langle \, {\bf
u}'_{2}\, m'_{s_{1}} \, m'_{s_{3}} \,| F(Q^{2})\ {\sigma_{1}\cdot
\widehat{{\bf Q}}\over Q^{2}+m_{\pi}^{2}} \ Q^{2} \ (\sigma_{3}\cdot
\widehat{{\bf Q}})^{2} | \, {\bf u}''_{2}\, m''_{s_{1}} \,
m''_{s_{3}} \, \rangle_{1} \nonumber \\ && =\, _{1}\langle \, {\bf
u}'_{2}\, m'_{s_{1}} \,|V_{31}^{(B)} | \, {\bf u}''_{2}\,
m''_{s_{1}} \, \rangle_{1} \ {\O} _{m'_{s_{3}} m''_{s_{3}}} ^{
\widehat{({\bf u}'_{2}-{\bf u}''_{2})} } \ ^{2},
\end{eqnarray}

The matrix elements of $V_{23}^{i}$ can be evaluated by following
the same algorithm as above. So we obtain:
\begin{eqnarray}
\label{W_23^i}  _{2}\langle \, {\bf u}'''\, \alpha''' \,
|V_{23}^{(i)}| \, {\bf u}''''\, \alpha'''' \,\rangle_{2} &=&
 \delta^{3}({\bf u}'''_{1}-{\bf u}''''_{1}) \ \delta^{3}({\bf
u}'''_{3}-{\bf u}''''_{3}) \ \delta_{\alpha'''_{T} \alpha''''_{T}} \
\nonumber \\ &\times& _{2}\langle \, {\bf u}'''_{2}\, \alpha'''_{S}
\, |V_{23}^{(i)}| \, {\bf u}''''_{2}\, \alpha''''_{S} \,\rangle_{2},
\end{eqnarray}
where
\begin{eqnarray}
\label{W_23^ABC} _{2}\langle \, {\bf u}'''_{2}\, \alpha'''_{S}
\,|V_{23}^{(A, B, C)}| \, {\bf u}''''_{2}\, \alpha''''_{S} \,
\rangle_{2} &=& \sum_{\gamma'''_{S}} \sum_{\gamma''''_{S}}
g_{\alpha'''_{S} \gamma'''_{S}}^{2} \ g_{\alpha''''_{S}
\gamma''''_{S}}^{2} \ \delta_{m'''_{s_{1}} m''''_{s_{1}}}
\delta_{m'''_{s_{3}}
m''''_{s_{3}}} \delta_{m'''_{s_{4}} m''''_{s_{4}}} \nonumber \\
&\times& \ _{2}\langle \, {\bf u}'''_{2}\, m'''_{s_{2}}
\,|V_{23}^{(A, B, C)}| \, {\bf u}''''_{2}\, m''''_{s_{2}} \,
\rangle_{2},
\end{eqnarray}
\begin{eqnarray}
\label{W_23^D} _{2}\langle \, {\bf u}'''_{2}\, \alpha'''_{S}
\,|V_{23}^{(D)}| \, {\bf u}''''_{2}\, \alpha''''_{S} \, \rangle_{2}
&=& \sum_{\gamma'''_{S}} \sum_{\gamma''''_{S}} g_{\alpha'''_{S}
\gamma'''_{S}}^{2} \ g_{\alpha''''_{S} \gamma''''_{S}}^{2} \
\delta_{m'''_{s_{1}} m''''_{s_{1}}}
\delta_{m'''_{s_{4}} m''''_{s_{4}}} \nonumber \\
&\times& \ _{2}\langle \, {\bf u}'''_{2}\, m'''_{s_{2}} \,
m'''_{s_{3}} \,|V_{23}^{(D)}| \, {\bf u}''''_{2}\, m''''_{s_{2}} \,
m''''_{s_{3}} \, \rangle_{2}.
\end{eqnarray}

The $A$-, $B$- and $C$-terms which have been considered in the right
side of Eq. (\ref{W_23^ABC}) can be evaluated as:
\begin{eqnarray}
\label{W_31^A}
 _{2}\langle \, {\bf u}'''_{2}\, m'''_{s_{2}} \,|V_{23}^{(A)} | \, {\bf
u}''''_{2}\, m''''_{s_{2}} \, \rangle_{2} &=& \ _{2}\langle \, {\bf
u}'''_{2}\, m'''_{s_{2}} \,|F(Q'^{2})\ {\sigma_{2}\cdot
\widehat{{\bf Q}}'\over Q'^{2}+m_{\pi}^{2}} \ Q'| \, {\bf
u}''''_{2}\, m''''_{s_{2}} \, \rangle_{2} \nonumber \\ &=&
\frac{F(({\bf u}''''_{2}-{\bf u}'''_{2})^{2})}{({\bf u}''''_{2}-{\bf
u}'''_{2})^{2}+m_{\pi}^{2}} \ |{\bf u}''''_{2}-{\bf u}'''_{2}| \
{\O} _{m'''_{s_{2}} m''''_{s_{2}}} ^{ \widehat{({\bf u}''''_{2}-{\bf
u}'''_{2})} },
\end{eqnarray}
\begin{eqnarray}
\label{W_31^B}
 _{2}\langle \, {\bf u}'''_{2}\, m'''_{s_{2}} \,|V_{23}^{(B)} | \, {\bf
u}''''_{2}\, m''''_{s_{2}} \, \rangle_{2} &=& \ _{2}\langle \, {\bf
u}'''_{2}\, m'''_{s_{2}} \,|F(Q'^{2})\ {\sigma_{2}\cdot
\widehat{{\bf Q}}'\over Q'^{2}+m_{\pi}^{2}} \  Q'^{2} | \, {\bf
u}''''_{2}\, m''''_{s_{2}} \, \rangle_{2} \nonumber \\ &=&
\frac{F(({\bf u}''''_{2}-{\bf u}'''_{2})^{2})}{({\bf u}''''_{2}-{\bf
u}'''_{2})^{2}+m_{\pi}^{2}} \ |{\bf u}''''_{2}-{\bf u}'''_{2}|^{2} \
{\O} _{m'''_{s_{2}} m''''_{s_{2}}} ^{ \widehat{({\bf u}''''_{2}-{\bf
u}'''_{2})} },
\end{eqnarray}
\begin{eqnarray}
\label{W_31^C}
 _{2}\langle \, {\bf u}'''_{2}\, m'''_{s_{2}} \,|V_{23}^{(C)} | \, {\bf
u}''''_{2}\, m''''_{s_{2}} \, \rangle_{2}&=& \ _{2}\langle \, {\bf
u}'''_{2}\, m'''_{s_{2}} \,| F(Q'^{2})\ {\sigma_{2}\cdot
\widehat{{\bf Q}}'\over Q'^{2}+m_{\pi}^{2}} \ Q'^{3}| \, {\bf
u}''''_{2}\, m''''_{s_{2}} \, \rangle_{2} \nonumber \\ &=&
\frac{F(({\bf u}''''_{2}-{\bf u}'''_{2})^{2})}{({\bf u}''''_{2}-{\bf
u}'''_{2})^{2}+m_{\pi}^{2}} \ |{\bf u}''''_{2}-{\bf u}'''_{2}|^{3} \
{\O} _{m'''_{s_{2}} m''''_{s_{2}}} ^{ \widehat{({\bf u}''''_{2}-{\bf
u}'''_{2})} },
\end{eqnarray}
also the different parts of the $D$-term which have been considered
in the right side of Eq. (\ref{W_23^D}) can be evaluated in the
following:
\begin{eqnarray}
\label{W_31^D_1_5}
 _{2}\langle \, {\bf u}'''_{2}\, m'''_{s_{2}}
\, m'''_{s_{3}} \,|V_{23}^{(D)} \ ^{1^{th}}| \, {\bf u}''''_{2}\,
m''''_{s_{1}} \, m''''_{s_{3}} \, \rangle_{2} &=& \  _{2}\langle \,
{\bf u}'''_{2}\, m'''_{s_{2}} \, m'''_{s_{3}} \,|V_{23}^{(D)} \
^{5^{th}}| \, {\bf u}''''_{2}\, m''''_{s_{1}} \, m''''_{s_{3}} \,
\rangle_{2} \nonumber \\ &=& \delta _{m'''_{s_{3}} m''''_{s_{3}}} \
_{2}\langle \, {\bf u}'''_{2}\, m'''_{s_{2}} \,|V_{23}^{(B)} | \,
{\bf u}''''_{2}\, m''''_{s_{1}} \, \rangle_{2}, \nonumber \\
\end{eqnarray}
\begin{eqnarray}
\label{W_31^D_2_4} && _{2}\langle \, {\bf u}'''_{2}\, m'''_{s_{2}}
\, m'''_{s_{3}} \,|V_{23}^{(D)} \ ^{2^{th}}| \, {\bf u}''''_{2}\,
m''''_{s_{1}} \, m''''_{s_{3}} \, \rangle_{2} \nonumber \\ && =  \,
_{2}\langle \, {\bf u}'''_{2}\, m'''_{s_{2}} \, m'''_{s_{3}}
\,|V_{23}^{(D)} \ ^{4^{th}}| \, {\bf u}''''_{2}\, m''''_{s_{1}} \,
m''''_{s_{3}} \, \rangle_{2} \nonumber \\ && = \, _{2}\langle \,
{\bf u}'''_{2}\, m'''_{s_{2}} \,| F(Q'^{2})\ { \sigma_{2}\cdot
\widehat{{\bf Q}}' \over Q'^{2}+m_{\pi}^{2}} \ Q'^{2} \
(\sigma_{3}\cdot \widehat{{\bf Q}}')^{2} | \, {\bf u}''''_{2}\,
m''''_{s_{2}} \, \rangle_{2} \nonumber \\ && = \, _{2}\langle \,
{\bf u}'''_{2}\, m'''_{s_{2}} \,|V_{23}^{(B)} | \, {\bf u}''''_{2}\,
m''''_{s_{1}} \, \rangle_{2} \ {\O} _{m'''_{s_{2}} m''''_{s_{2}}} ^{
\widehat{({\bf u}''''_{2}-{\bf u}'''_{2})} } \ ^{2},
\end{eqnarray}
\begin{eqnarray}
\label{W_31^D_3} && _{2}\langle \, {\bf u}'''_{2}\, m'''_{s_{2}} \,
m'''_{s_{3}} \,|V_{23}^{(D)} \ ^{3^{th}}| \, {\bf u}''''_{2}\,
m''''_{s_{1}} \, m''''_{s_{3}} \, \rangle_{2} \nonumber \\ && =  \,
_{2}\langle \, {\bf u}'''_{2}\, m'''_{s_{2}} \,| F(Q'^{2})\ {
\sigma_{2}\cdot \widehat{{\bf Q}}' \over Q'^{2}+m_{\pi}^{2}} \
Q'^{2} \ \sigma_{3}\cdot \widehat{{\bf Q}}' | \, {\bf u}''''_{2}\,
m''''_{s_{2}} \, \rangle_{2} \nonumber \\ && =  \, _{2}\langle \,
{\bf u}'''_{2}\, m'''_{s_{2}} \,|V_{23}^{(B)} | \, {\bf u}''''_{2}\,
m''''_{s_{1}} \, \rangle_{2} \ {\O} _{m'''_{s_{2}} m''''_{s_{2}}} ^{
\widehat{({\bf u}''''_{2}-{\bf u}'''_{2})} }.
\end{eqnarray}

\section{Parity and Time Reversal Properties of the Total 4N Wave Function}
\label{appendix:Parity_Time-reversal}

In our formulation of 4N bound state we have not yet used the
properties of total wave function under the parity and time reversal
invariance. In this section we discuss about these properties.

The parity invariance would mean:
\begin{eqnarray}
\label{parity}
 \langle \, {\bf u} \,\,  \alpha  \, | \Psi \rangle &=&
 \langle \, {\bf u} \,\,  \alpha  \, |P P |\Psi \rangle \nonumber \\ &=&
 \langle \, -{\bf u} \,\,  \alpha  \, |P |\Psi \rangle , \quad  P |\Psi \rangle=  |\Psi \rangle \nonumber \\ &=&
 \langle \, -{\bf u} \,\,  \alpha  \, |\Psi \rangle
\end{eqnarray}

The parity invariance or equivalently the symmetry property of the
total wave function $| \Psi \rangle$ under the exchange of the
vector sets $|{\bf u}\rangle$ to $|-{\bf u}\rangle$ can be used in
the numerical solution of the coupled equations (\ref{YCs_u_v}) to
reduce the dimension of the 4N problem. On the other hand this
symmetry property can be used explicitly to cut down the size of the
Yakubovsky components and thus save the time and memory when
computing the coupled three dimensional integral equations.

The time reversal invariance might be more interesting. The total
wave function can be written as:
\begin{eqnarray}
\label{WF_alpha_gamma}
 \langle \, {\bf u} \,\,  \alpha  \, | \Psi \rangle =
 \sum_{\gamma}  \langle   \alpha   | \gamma \rangle \,
 \langle \, {\bf u} \,\,  \gamma  \, |\Psi \rangle =
 \sum_{\gamma}  g_{\alpha \, \gamma} \,
 \langle \, {\bf u} \,\,  \gamma  \, |\Psi \rangle,
\end{eqnarray}
where
\begin{eqnarray}
\label{time_reversal}
 \langle \, {\bf u} \,\,  \gamma  \, | \Psi \rangle &=&
 \langle \, {\bf u} \,\,  \gamma  \, |T \, T |\Psi \rangle \nonumber \\
 &=& \prod_{i=1}^{4} \left(- \right)^{(\frac{1}{2}-m_{s_{i}})} \, \left(- \right)^{(\frac{1}{2}-m_{t_{i}})}
 \langle \, -{\bf u} \,\,  -\gamma  \, |T |\Psi \rangle  \nonumber \\
 &=& \prod_{i=1}^{4} \left(- \right)^{(1-m_{s_{i}}-m_{t_{i}})}  \langle \, -{\bf u} \,\,  -\gamma  \, |T |\Psi
 \rangle.
\end{eqnarray}

By using the time reversal invariance of the total wave function,
which is applicable for the ground state of $^{4}$He with the total
angular momentum $J=0$, i.e. $T |\Psi \rangle=  |\Psi \rangle$, as
well as the parity invariance, Eq. (\ref{parity}), we can rewrite
Eq. (\ref{time_reversal}) as:
\begin{eqnarray}
\label{time_reversal_final}
 \langle \, {\bf u} \,\,  \gamma  \, | \Psi \rangle = \prod_{i=1}^{4} \left(- \right)^{(1-m_{s_{i}}-m_{t_{i}})}
 \langle \, {\bf u} \,\,  -\gamma  \,  |\Psi \rangle
\end{eqnarray}

The time reversal, and the parity invariance, as indicated in Eq.
(\ref{time_reversal_final}) can be also used to cut down the size of
the total wave function, by reducing the dimension of the
spin-isospin parts of total wave function, which will be valuable in
the calculations of 3NFs.

\end{document}